\documentstyle[preprint,aps]{revtex}

\begin{document}
\title{Quantum phase transitions and thermodynamic properties in highly anisotropic
magnets}
\author{V.Yu.Irkhin$^{*}$ and A.A.Katanin}
\address{620219, Institute of Metal Physics, Ekaterinburg, Russia}
\maketitle

\begin{abstract}
The systems exhibiting quantum phase transitions (QPT) are investigated
within the Ising model in the transverse field and Heisenberg model with
easy-plane single-site anisotropy. Near QPT a correspondence between
parameters of these models and of quantum $\phi ^4$ model is established. A
scaling analysis is performed for the ground-state properties. The influence
of the external longitudinal magnetic field on the ground-state properties
is investigated, and the corresponding magnetic susceptibility is
calculated. Finite-temperature properties are considered with the use of the
scaling analysis for the effective classical model proposed by Sachdev.
Analytical results for the ordering temperature and temperature dependences
of the magnetization and energy gap are obtained in the case of a small
ground-state moment. The forms of dependences of observable quantities on
the bare splitting (or magnetic field) and renormalized splitting turn out
to be different. A comparison with numerical calculations and experimental
data on systems demonstrating magnetic and structural transitions (e.g.,
into singlet state) is performed.
\end{abstract}


\section{Introduction}

The interest in quantum models of anisotropic spin and pseudospin systems is
connected with that they describe miscellaneous magnetic and structural
transitions. Examples of such transitions are transitions into singlet
magnetic state in TbSb, Pr, Pr$_3$Tl (see Ref. \cite{CooperBook} and
references therein), NiSi$_2$F$_6$ (see, e.g., Ref. \cite{FSiNi}), and
orientational and metamagnetic phase transitions under magnetic field\cite
{Coq,Levitin}.

The simplest model for the systems demonstrating a ground-state quantum
phase transition (QPT) is the Ising model in the transverse field. This
model is convenient for description of structural transitions in quantum
crystals \cite{Samara,Blinc,Gehring}. It can be also applied to describe
magnetic systems where both lowest and next energy levels are singlets. A
more complicated first-principle model for spin systems in a strong crystal
field is the Heisenberg model with an easy-plane single-site anisotropy; it
is applicable in the case where next-to-lowest energy level is doublet. As
well as transverse-field Ising model, this model also demonstrates QPT (the
ground-state magnetization vanishes with increasing the anisotropy
parameter).

A number of approximate methods were applied to study the transverse-field
Ising model \cite{MF,Bosons,Grover,RPA-TSCA,HTSE,GSPT,c1/z,Shender,Numer} at
$d>1$ and the Heisenberg model with easy-axis anisotropy \cite
{QPT-Epl,Fl-Epl,HTSE-Pl,Onufr,Valkov}. However, all these methods (except
for numerical ones) are applicable only not too close to QPT. In particular,
they lead to Gaussian values of the QPT critical exponents. Thus analytical
consideration of the ground-state and finite-temperature properties in the
vicinity of QPT is still an open problem.

The case, where the system is close to QPT, is characterized by a small
ground-state moment and low transition temperature. Such a situation is
reminiscent of weak itinerant magnets\cite{Moriya}. An analysis of the
ground-state QPT was performed in Refs.\cite{Hertz,Millis}. It was shown
that the upper critical dimensionality for such transitions is $d_c^{+}=4-z$
with $z$ being the dynamical critical exponent. This conclusion has a
general character. In the present paper we consider only systems with $z=1,$
which holds for the transverse-field Ising model and anisotropy-induced QPT
in the Heisenberg model. (Note that this is not the case for the QPT induced
by magnetic field in degenerate systems with $n>1$ component order parameter
since here $z=2,$ see Ref. \cite{XYh}). Thus for $d\geq 3$ the critical
exponents are the Gaussian ones, while for $d<3$ they deviate from the
corresponding mean-field values and can be calculated with the use of the $%
3-\varepsilon $ expansion. For the critical dimensionality $d=3$ the ground
state properties contain logarithmic corrections.

Sachdev\cite{Sachdev} proposed a three-stage method of treating
finite-temperature properties of the systems near QPT. At the first stage,
ground-state renormalizations are performed. At the second stage, the
non-zero Matsubara frequencies are integrated out to obtain an effective
classical action. Finally, perturbation theory for the effective classical
model is applied. This method ensures correct analytical properties of the
resulting theory. While ground-state renormalizations are non-universal,
finite-temperature properties, being expressed through quantum-renormalized
ground-state parameters, turn out to be universal.

The approach of Ref. \cite{Sachdev} is based on a continuum model, namely,
the quantum $\phi ^4$ model. This model is sufficient to express
finite-temperature properties near QPT through the non-universal
ground-state properties, but insufficient to obtain correct results for the
latter. A convenient method to consider the lattice spin systems near their
critical dimensionality is the expansion in the formal quasiclassical
parameter $1/S$. Its applicability is connected with the fact that near $d_c$
the effective interaction of spin waves is small (except for a narrow
critical region where the $\varepsilon $-expansion can be easily developed
to correct the description of the critical behavior). For the Heisenberg
model such a situation occurs for temperature transition near the lower
critical dimensionality $d_c^{-}=2$. This provides success of the
renormalization-group (RG) approach for the description of thermodynamics of
$d=2$ (Ref.\cite{Chakraverty}) and $d=2+\varepsilon $ (Refs. \cite
{Brezin,Chakraverty}) Heisenberg magnets, and also quasi-2D and anisotropic
2D magnets \cite{OurRG} not too close to $T_c$.

For QPT in highly-anisotropic spin systems the $1/S$-expansion works well
near the {\it upper }critical dimensionality $d_c^{+}=3.$ In this case there
are excitations, which are almost gapless near QPT\ (they are analogous to
the spin-wave excitations in Heisenberg magnets). Besides that, for the
ordered degenerate systems (with $n\geq 2$) there are always Goldstone modes
with zero energy gap and the $1/S$-expansion becomes applicable at arbitrary
anisotropy below its critical value. The situation is more complicated for
finite temperatures, since close to temperature transition the system
behaves as a corresponding classical magnet and therefore the picture of
excitation spectrum differs from that at $T=0.$

The aim of the present paper is to apply the above-discussed concepts for
calculating ground-state and finite-temperature properties of
transverse-field Ising model ($n=1$) and Heisenberg model with strong
easy-plane anisotropy ($n=2$). To this end we apply perturbation theory
(which is in fact an expansion in $1/S$) to the original lattice models (not
to their continuum analogs), which enables us to calculate non-universal
ground-state quantum renormalizations. After that we combine perturbation
results for short-wave fluctuations with the results of the $3-\varepsilon $
RG approach for the long-wave fluctuations to correct the results of
perturbation theory. Finally, we consider finite-temperature properties
within RG approach for the effective continuum classical model.

The plan of the paper is as follows. In Sect.2 we discuss the Ising model in
the transverse field. We consider corresponding mean-field results,
construct the perturbation theory in $1/S$ and apply a scaling approach to
investigate ground-state and thermodynamic properties, in particular the
influence of external magnetic field. In Sect.3 the Heisenberg model with
easy-plane anisotropy is considered in a similar way. In Sect.4 we discuss
the results obtained and compare them with experimental data on systems
exhibiting structural and magnetic transitions. Some details of calculations
are presented in Appendices.

\section{Transverse-field Ising model}

\subsection{The formulation of the model and the mean-field approximation}

We consider the Hamiltonian of the Ising model in the transverse field $%
\Omega $
\begin{equation}
{\cal H}=-\frac I2\sum_{\langle ij\rangle }S_i^xS_j^x-\Omega \sum_iS_i^z
\label{ITF}
\end{equation}
where $I$ is the exchange parameter. This model can describe singlet
magnetic systems. A derivation of such a model for Heisenberg magnets with
strong single-site anisotropy is presented in Appendix A. The model (\ref
{ITF}) describes also structural transition in quantum crystals (cooperative
Jahn-Teller effect, see Ref. \cite{Gehring}) where two lowest energy levels
are singlets. In this case $I=\Delta _2,$ $\Omega =\Delta _1,$ where $\Delta
_{1,2}$ is the energy-level splitting at $T=0$ and $T>T_c$ respectively. For
further purposes it will be useful to consider the model (\ref{ITF}) for
arbitrary values of (pseudo-) spin $S.$

At $\Omega =0$ the model (\ref{ITF}) coincides with the Ising model and thus
the order parameter $\overline{S}\equiv \langle S^x\rangle =S$ in the ground
state. With increasing $\Omega $, the model (\ref{ITF}) demonstrates a
quantum phase transition where $\overline{S}$ vanishes. The one-dimensional $%
S=1/2$ transverse-field Ising model in the ground state can be solved
rigorously \cite{one-dim}. In particular, it can be reduced to the
two-dimensional Ising problem at finite temperatures \cite{Suzuki}, so that
critical exponents for both the phase transitions coincide. The
transverse-field Ising model with $d>1$ requires approximate methods.

The mean-field theory \cite{Blinc,MF} yields the critical field $\Omega
_0\equiv I_0S,$ and the equation for the order parameter at $\Omega <\Omega
_0$ reads
\begin{equation}
\frac{\Omega _0}{H_e}B_S\left( H_e/T\right) =1
\end{equation}
where
\begin{eqnarray}
B_S(x) &=&(1+1/2S)\coth (1+1/2S)x-(1/2S)\coth (x/2S),  \nonumber \\
B_{1/2}(x) &=&(1/2)\tanh (x/2)
\end{eqnarray}
is the Brillouin function, $H_e=(\Omega ^2+\Omega _0^2\overline{S}%
^2/S^2)^{1/2},$ $I_0=2Id.$ Owing to the field $\Omega ,$ the value of $%
\langle S^z\rangle $ is finite in both ordered and disordered phase and
reads 
\begin{equation}
\langle S^z\rangle =\frac{\Omega S}{H_e}B_S\left( H_e/T\right) .
\end{equation}
It should be noted that at $\Omega <\Omega _0$ we have simply $\langle
S^z\rangle =S\Omega /\Omega _0.$ The critical temperature where $\overline{S}
$ vanishes is determined for the physically important case $S=1/2$ by 
\begin{equation}
T_c^{MF}=\frac \Omega {2\tanh ^{-1}(\Omega /\Omega _0)}\simeq \frac \Omega {%
\ln [2/(1-\Omega /\Omega _0)]}  \label{TcMF}
\end{equation}
(the last approximation is valid for $1-\Omega /\Omega _0\ll 1$). Thus the
mean-field theory predicts a very weak inverse-logarithmic dependence for
the critical temperature near QPT in arbitrary dimensionality. This
contradicts to the results of the scaling approach \cite{Millis,Sachdev}
both above and below the upper critical dimensionality $d_c^{+}=3$.

To improve the mean-field theory, one has to take into account the
collective excitations which are analogous to spin-wave excitations in
Heisenberg magnets. The spectrum of these excitations in the random-phase
approximation has the form \cite{Blinc} 
\begin{equation}
E_{{\bf q}}^2=\Omega [\Omega -I_{{\bf q}}\langle S^z\rangle ]+I_0^2\overline{%
S}^2  \label{EqRPA}
\end{equation}
in both ordered and disordered phases. Near QPT (at $\overline{S}(T=0)\ll 1$%
), these excitations become almost gapless and give dominant contributions
to physical properties.

The result of account of the collective excitations to first order in $1/%
{\cal R}$ (where ${\cal R}$ is the radius of exchange interaction) \cite
{Shender} for $d=3$ reads 
\begin{equation}
T_c\sim {\cal R}^{3/2}\sqrt{1-\Omega /\Omega _0}.  \label{TcR}
\end{equation}
This has a correct square-root behavior (see, e.g., \cite{Sachdev})$.$
However, the logarithmic corrections, that occur for $d=3,$ are not
reproduced by the result (\ref{TcR}). Besides that, the $1/{\cal R}$
expansion does not enable one to determine correctly the coefficient in (\ref
{TcR}) for not too large ${\cal R}$.

Another approach used in \cite{RPA-TSCA} is to consider the excitations (\ref
{EqRPA}) self-consistently within the random-phase approximation (RPA)
decoupling scheme for the sequence of equations of motion. Unlike the $1/%
{\cal R}$ expansion, this procedure gives a possibility to take into account
the reaction of the RPA\ excitation spectrum (\ref{EqRPA}) on the deviation
of the critical field from $\Omega _0$. However, corrections to mean-field
ground-state parameters turn out to be small enough, and at finite
temperatures the RPA\ magnetization shows a double-value behavior with
first-order temperature phase transition. Authors of Ref. \cite{RPA-TSCA}
consider also a generalization of RPA, the two-site self-consistent
approximation (TSCA) which gives a possibility to include partially
correlation effects. This approximation gives more satisfactory results than
RPA. However, it predicts first-order character not only for the temperature
transition, but also for QPT.

One should mention also the papers\cite{HTSE,GSPT} where high-temperature
series expansions (HTSE) and ground-state perturbation theory (GSPT)\ were
used. Although these expansions gives consistent results for the critical
field, their applicability near QPT is questionable. Recently some results
of GSPT and HTSE have been confirmed by numerical correlated-basis-function
analysis \cite{Numer}.

Below we use the $1/S$ expansion to treat ground-state and finite
temperature properties of the transverse-field Ising model. Unlike the $1/%
{\cal R}$ expansion, it takes into account the ``spin-wave'' excitations
already in zeroth order of perturbation theory. Contrary to RPA \cite
{RPA-TSCA}, this is a systematic expansion, and therefore scaling
corrections can be easily calculated. It should be also noted that the $1/S$
expansion differs from the ground-state perturbation theory used in Ref. 
\cite{GSPT} where the expansions in powers of $\Omega /I$ and $I/\Omega $
are used for the ordered and disordered phases respectively. Indeed, the $%
1/S $ expansion treats both the terms in the Hamiltonian (\ref{ITF}) on
equal footing and thus yields physically correct results already in the
first order in $1/S.$

\subsection{\label{Sec2}Ground-state properties within the $1/S$
perturbation theory}

To construct perturbation expansion in a convenient form we use the spin
coherent state approach\cite{Klauder}. The partition function is presented
in terms of a path integral, 
\begin{equation}
{\cal Z}=\int D{\bbox \pi }\exp [-({\cal S}_{\text{dyn}}{\cal +S}_{\text{st}%
})],  \label{Zp}
\end{equation}
where 
\begin{eqnarray}
{\cal S}_{\text{dyn}} &=&iS\sum_i\int\limits_0^{1/T}d\tau \left( 1-\cos
\vartheta _i\right) \frac{\partial \varphi _i}{\partial \tau }  \nonumber \\
{\cal S}_{\text{st}} &=&-\int\limits_0^{1/T}d\tau \left[ \frac{IS^2}2%
\sum_{\langle ij\rangle }\pi _{xi}\pi _{xj}+\Omega S\sum_i\pi _{zi}\right] 
\end{eqnarray}
are static and dynamic parts of the action, ${\bbox \pi }_i=\{\pi _{xi},\pi
_{yi},\pi _{zi}\}$ is three-component vector field with ${\bbox \pi }%
_i^2=1+1/S$, $\vartheta _i$ and $\varphi _i$ are the polar and azimuthal
angles of ${\bbox \pi }_i$ in an arbitrarily chosen coordinate system (which
does not need to coincide with the $\pi _x$-$\pi _y$-$\pi _z$ coordinate
system). Further we additionally rotate the coordinate system through the
angle $\theta $ determined by 
\begin{equation}
\sin \theta =\langle \pi _x\rangle /\langle |{\bbox \pi }|\rangle 
\end{equation}
around $\pi _y$-axis (in the disordered phase $\theta =0$ and the rotated
coordinate system coincides with the original one). Then $\langle \widetilde{%
\pi }_x\rangle =0$ in the new coordinate system.

The calculation of two-point vertex function $\Gamma ({\bf q},\omega )$ of
the fields $\widetilde{\pi }_x,\,\widetilde{\pi }_y$ (the tilde sign is
referred to rotated coordinate system), which is connected with matrix
Green's function $G$ of these fields by the relation $\Gamma ({\bf q},\omega
)=G^{-1}({\bf q},\omega )$, is performed for both ordered and disordered
phases in Appendix B and yields to first order in $1/S$ the result 
\begin{equation}
\Gamma _{\pm }({\bf q},\omega _n)=\left( 
\begin{array}{cc}
S^2(I_0-I_{{\bf q}}+I_0\Delta _{\pm }^2) & iS\omega _n\left(
w_S+X_0/2+Y_0/2\right) \\ 
iS\omega _n\left( w_S+X_0/2+Y_0/2\right) & I_0S^2D_{\pm }
\end{array}
\right)  \label{Gamma}
\end{equation}
where $w_S=(1+1/2S)^{-1},$ 
\begin{equation}
X_0=\frac 1{2S}\sum_{{\bf q}}\frac 1{\sqrt{1-I_{{\bf q}}/I_0}},\;Y_0=\frac 1{%
2S}\sum_{{\bf q}}\sqrt{1-I_{{\bf q}}/I_0},
\end{equation}
$\Delta _{\pm }$ and $D_{\pm }$ are the dimensionless temperature-dependent
energy gap and the renormalization factor for the exchange parameter in the
disordered and ordered phases respectively, their concrete expressions being
specified below. The matrix static uniform spin susceptibility in the
rotated coordinate system is expressed through $\Gamma $ as 
\begin{equation}
\widetilde{\chi }^{ij}=S^2\Gamma _{ij}^{-1}(0,0)  \label{Hij}
\end{equation}
where $i,j=x,y.$ The renormalized spectrum of ``spin-wave'' excitations is
determined by the condition $\det \Gamma ({\bf q,-}iE_{{\bf q}})=0$ and to
first order in $1/S$ has the form 
\begin{equation}
\widetilde{E}_{{\bf q}}=S\left[ 1+1/2S-(X_0+Y_0)/2\right] \sqrt{I_0D_{\pm
}(I_0-I_{{\bf q}}+I_0\Delta _{\pm }^2)}.  \label{Eqr}
\end{equation}
The quantum-renormalized critical field $\Omega _c$ is given by 
\begin{equation}
\frac{\Omega _c}{\Omega _0}=1+\frac 1{2S}-\frac 1{4S}\sum_{{\bf q}}\frac{%
2I_0+I_{{\bf q}}}{\sqrt{I_0(I_0-I_{{\bf q}})}}.  \label{Wc}
\end{equation}
Last two terms in this expression yield the first-order $1/S$-correction to
mean-field value of $\Omega _c.$ For $S=1/2$ numerical calculation of
integral in (\ref{Wc}) yields the result $\Omega _c=2.44I$ in the 3D case
and $\Omega _c=1.10I$ in the 2D case. Thus the critical field is strongly
renormalized by quantum fluctuations both in 3D and 2D cases. The critical
field values obtained are considerably smaller than the corresponding RPA
results\cite{RPA-TSCA}, $\Omega _c=2.88I$ and $\Omega _c=1.83I$ and somewhat
smaller than those obtained by HTSE \cite{HTSE} and GSPT \cite{GSPT}, $%
\Omega _c=2.58I$ and $\Omega _c=1.54I.$ This demonstrates that considered
first-order $1/S$ perturbation theory overestimates effects of quantum
fluctuations (especially in the 2D case), but treats these fluctuations more
correctly than RPA.

In the disordered phase with $\langle \pi _x\rangle =0$ ($\Omega >\Omega _c$%
) the expressions for the ground-state energy gap and factor $D_{+}$ has the
form 
\begin{eqnarray}
\Delta _{+}^2(t_{+},0) &=&\frac{t_{+}}{1-t_{+}}\left( 1-X_0^{\prime }\right)
\left[ 1+A_{+}(t_{+})\right]  \nonumber \\
\ A_{+}(t) &=&\frac 1{4St}\sum_{{\bf q}}\left\{ \frac{2I_0+I_{{\bf q}}(1+t)}{%
\sqrt{I_0[I_0-I_{{\bf q}}(1-t)]}}-\frac{2I_0+I_{{\bf q}}}{\sqrt{I_0(I_0-I_{%
{\bf q}})}}\right\}  \label{DeltaP}
\end{eqnarray}
and 
\begin{equation}
D_{+}(t_{+})=\frac{1+Y_0-X_0}{1-t_{+}}\left[ 1+t_{+}A_{+}(t_{+})\right]
\label{DP}
\end{equation}
where 
\begin{equation}
t_{+}=1-\Omega _c/\Omega .
\end{equation}
and 
\[
X_0^{\prime }=\frac 1{2S}\sum_{{\bf q}}\frac{I_{{\bf q}}}{\sqrt{I_0(I_0-I_{%
{\bf q}})}} 
\]
In the ordered phase ($\Omega <\Omega _c$) we obtain 
\begin{eqnarray}
\Delta _{-}^2(t_{-},0) &=&t_{-}\left( 1-X_0^{\prime }\right) \left[
1+A_{-}(t_{-})\right] ,  \nonumber \\
A_{-}(t) &=&-2(1-t)A_{+}(t)  \nonumber \\
&&\ \ \ -\frac{1-t}{8S}\sum_{{\bf q}}\frac{[2I_0+I_{{\bf q}}(1+t)]^2}{%
I_0^{1/2}[I_0-(1-t)I_{{\bf q}}]^{3/2}}  \label{DeltaM}
\end{eqnarray}
and 
\begin{equation}
D_{-}=1+Y_0-X_0  \label{DM}
\end{equation}
where 
\begin{equation}
t_{-}=1-(\Omega /\Omega _c)^2.
\end{equation}

Consider now the observable quantities. The expression for the order
parameter $\overline{S}(t_{-},T)\equiv S\langle \pi _x\rangle $ at $T=0,$ $%
\Omega <\Omega _c$ reads 
\begin{eqnarray}
\overline{S}(t_{-},0) &=&St_{-}^{1/2}\left[ 1+B(t_{-})\right] ^{1/2}\left[
1+1/2S-(X_0+Y_0)/2\right] ,  \label{Sl} \\
B(t) &=&-2(1-t)A_{+}(t)  \nonumber
\end{eqnarray}
For the longitudinal susceptibility we have 
\begin{equation}
\chi ^{xx}=\cos ^2\theta \widetilde{\chi }^{xx}+\sin \theta \cos \theta
\left( \widetilde{\chi }^{xz}+\widetilde{\chi }^{zx}\right) +\sin ^2\theta 
\widetilde{\chi }^{zz}  \label{HiOrd}
\end{equation}
where the tilde sign is referred to susceptibilities in the rotated
coordinate system (recall that for the disordered phase $\theta =0$). For
the ordered phase the first summand in (\ref{HiOrd})\ gives dominant
contribution near QPT, and using the relation (\ref{Hij}) yields in both the
ordered phase near QPT and disordered phase the expression for the
ground-state spin susceptibility through the gap in the excitation spectrum 
\begin{equation}
\chi ^{xx}=\frac 1{I_0\Delta _{\pm }^2(t_{\pm },0)}.  \label{HiOrd1}
\end{equation}

In the limiting case of zero transverse field we have the trivial result $%
\overline{S}=S,$ and the excitation spectrum reduces to its mean-field form, 
$\widetilde{E}_{{\bf q}}=\Omega _0.$ At very large $\Omega \gg \Omega _c$ we
reproduce again the mean-field result $\widetilde{E}_{{\bf q}}=\Omega $.
This is a consequence of the fact that in both the limits $\Omega =0$ and $%
\Omega \rightarrow \infty $ quantum fluctuations are absent. Thus the $1/S$%
-expansion gives a possibility to describe the ground-state properties for
an arbitrary $\Omega \geq 0.$ However, as we shall see below, some
difficulties arise in the region $\Omega \approx \Omega _c\;$where the
quantum fluctuations are strong enough to modify considerably the results. A
more detailed consideration of this region will be performed in Sect.\ref
{Sec3a}.

\subsection{\label{Sec3}Influence of longitudinal magnetic field}

To consider the influence of the external magnetic field we add to the
Hamiltonian the term 
\begin{equation}
\Delta {\cal H}=-H\sum_iS_i^x.
\end{equation}
The longitudinal magnetic field results in the appearance of nonzero $%
\langle S_i^x\rangle $ at any $\Omega /I.$ The influence of both transverse
and external longitudinal fields is, of course, equivalent to applying one
effective field which has the value $(H^2+\Omega ^2)^{1/2}$ and makes the
angle $\arctan (H/\Omega )$ with the $\pi _x$-axis. However, it is useful to
consider these fields as two independent ones.

Performing the calculations which are similar to those of Sect.\ref{Sec2}
and Appendix B we obtain to first order in $1/S$ the equation for the angle $%
\theta $ of coordinate system rotation 
\begin{equation}
\Omega -\Omega _0r(\theta )\cos \theta -H\cot \theta =0,  \label{FEqA}
\end{equation}
where 
\begin{equation}
r(\theta )=1+\frac 1{2S}-\frac 1{4S}\sum_{{\bf q}}\frac{2(I_0+I_{{\bf q}%
})\varphi (\theta )-I_{{\bf q}}\cos ^2\theta }{\sqrt{I_0\varphi (\theta
)[I_0\varphi (\theta )-I_{{\bf q}}\cos ^2\theta ]}},
\end{equation}
with $\varphi (\theta )=\sin ^2\theta +(\Omega /\Omega _c)\cos \theta $. For
a general $\Omega /\Omega _c$ the solution of this equation is rather
cumbersome. However, near QPT (i.e. at $1-\Omega /\Omega _c\ll 1$), where
the angle $\theta $ is small, one can expand (\ref{FEqA}) in $\theta $ to
obtain 
\begin{equation}
\theta =\left\{ 
\begin{array}{cc}
\theta _0+H/2[I_0Sr(0)-\Omega ] & \theta _H\ll \theta _0 \\ 
\theta _H+2\delta r/3\theta _H & \theta _0\ll \theta _H\ll 1
\end{array}
\right.
\end{equation}
where $\theta _0=\sqrt{2(1-\Omega /\Omega _c)}$, $\theta _H=(2H/\Omega
_c)^{1/3}$\ and $\delta r=r(\theta _H)-r(0)$.

For the magnetization we derive 
\begin{equation}
\overline{S}=\left\{ 
\begin{array}{lc}
\overline{S}(H=0)+\chi ^{xx}H, & \theta _H^2\ll 1-\Omega /\Omega _c, \\ 
\left( 1+\frac 1{2S}-\frac{X_0+Y_0}2\right) \theta _H\left[ 1+B^{\prime
}(\theta _H^2)\right] , & 1-\Omega /\Omega _c\ll \theta _H^2\ll 1,
\end{array}
\right.  \label{FMagn}
\end{equation}
where $\chi ^{xx}$ is determined by (\ref{HiOrd1}), and 
\begin{equation}
B^{\prime }(\theta _H^2)=-\frac 1{6S\theta _H^2}\sum_{{\bf q}}\left[ \frac{%
2(I_0+I_{{\bf q}})(1+\theta _H^2/2)-I_{{\bf q}}(1-\theta _H^2)}{\sqrt{%
I_0[I_0(1+\theta _H^2)-I_{{\bf q}}(1-\theta _H^2/2)]}}-\frac{2I_0+I_{{\bf q}}%
}{\sqrt{I_0(I_0-I_{{\bf q}})}}\right] .
\end{equation}
The ground-state energy gap is given by 
\begin{equation}
\Delta _{-}^2\simeq \left\{ 
\begin{array}{lc}
\Delta _{-}^2(H=0), & \theta _H^2\ll 1-\Omega /\Omega _c, \\ 
\frac 32\theta _H^2\left[ 1+A^{\prime }(\theta _H^2)\right] , & 1-\Omega
/\Omega _c\ll \theta _H^2\ll 1,
\end{array}
\right.
\end{equation}
where 
\begin{equation}
A^{\prime }(\theta _H^2)=B^{\prime }(\theta _H^2)-\frac 1{12S}\sum_{{\bf q}}%
\frac{(I_0+2I_{{\bf q}})^2}{I_0^{1/2}[I_0(1+\theta _H^2/2)-I_{{\bf q}%
}(1-\theta _H^2)]^{3/2}}.
\end{equation}

The longitudinal susceptibility in the presence of magnetic field is still
determined by (\ref{HiOrd}), and again the first term gives main
contribution near QPT. Alternatively, the same result can be obtained by
direct differentiation of $\overline{S}$ [which is given by (\ref{FMagn})]\
with respect to $H.$

\subsection{\label{Sec3a}Ground-state renormalizations near QPT}

The results of $1/S$ expansion can be applied only at not too small $t_{\pm
} $. Indeed, at $d\leq 3$ the functions $A$ and $B$ contain terms which are
divergent at $t_{\pm }\rightarrow 0$ as $t_{\pm }^{(d-3)/2}$ (at $d=3,$
logarithmic divergences are present). The same situation takes place for the
functions $A^{\prime }$ and $B^{\prime }$ which are divergent as $\theta
_H^{d-3}$ at $\theta _H\rightarrow 0.$ Thus an $\varepsilon =3-d$ expansion
can be developed within the RG approach to treat these divergences more
correctly and to improve thereby the behavior of $\Delta _{\pm }$ and $%
\overline{S}$ near QPT. Further consideration of this section is related to
the critical region $|1-\Omega /\Omega _c|\ll 1.$ However, as it will be
clear below, the results can be extrapolated to arbitrary $\Omega ,$ since
in the limits $\Omega \ll \Omega _c$ and $\Omega \gg \Omega _c$ they are
smoothly joined with the results of the $1/S$ expansion of section \ref{Sec2}%
.

First we pick up the nonuniversal factors from $\overline{S},$ $\Delta _{\pm
}$ by introducing the quantities 
\begin{eqnarray}
\overline{S}_R(t,T) &=&\left[ 1+1/2S-(X_0+Y_0)/2\right] ^{-1}\overline{S}%
(t,T)/S  \nonumber \\
\Delta _{\pm R}(t,T) &=&\left( 1-X_0^{\prime }\right) ^{-1}\Delta _{\pm
}(t,T)
\end{eqnarray}
Consider the continuum limit of the above theory. The action ${\cal S}={\cal %
S}_{\text{dyn}}+{\cal S}_{\text{st}}$ in this limit takes the form 
\begin{eqnarray}
{\cal S}_{\text{cont}} &=&\frac 12\int d^dr\int\limits_0^{c/T}d\tau \left[
\,2i\widetilde{\pi }_x(\partial \widetilde{\pi }_y/\partial \tau )+%
\widetilde{\pi }_y^2+(\nabla \widetilde{\pi }_x)^2+m^2\widetilde{\pi }%
_x^2\right]  \nonumber \\
&&\ \ \ \ \ \ \ \ \ \ \ \ \ +\frac u{4!}\int d^dr\int\limits_0^{c/T}d\tau \,%
\widetilde{\pi }_x^4,  \label{Seff}
\end{eqnarray}
where the parameters $u$, $m^2$ and $c,$ determined in such a way, are given
by 
\begin{eqnarray}
u_{\text{cont}} &=&6d\frac{c_0}{IS^2}\zeta ,  \nonumber \\
c_{\text{cont}} &=&c_0,  \nonumber \\
m_{\text{cont}}^2 &=&2t_{+}d,  \label{pcont}
\end{eqnarray}
$c_0=(2d)^{1/2}IS$ being the bare spin-wave velocity,$\;\widetilde{\pi }%
_x^2=(IS^2/c_0)\pi _x^2,\;\widetilde{\pi }_y^2=(IS^2/c_0)\pi _y^2$ and the
factor $\zeta $\ ($\zeta =1-t_{-}$ in the ordered phase and $\zeta =1$ in
the disordered phase) is introduced to extend the region of applicability of
results obtained to arbitrary $t_{\pm }$. Note that the coefficients at
first three terms of the quadratic part of (\ref{Seff}) can be always chosen
equal to their values in (\ref{Seff}) by appropriate rescaling of $\pi
_{x,y} $ and $\tau .$ The model (\ref{Seff}) is completely equivalent to the
quantum $\phi ^4$ model. Indeed, integrating out the field $\pi _y$ we
obtain 
\begin{equation}
{\cal S}_{\text{cont}}=\frac 12\int d^dr\int\limits_0^{c/T}d\tau \left[
\,(\partial \widetilde{\pi }_x)^2+m^2\widetilde{\pi }_x^2\right] +\frac u{4!}%
\int d^dr\int\limits_0^{c/T}d\tau \,\widetilde{\pi }_x^4  \label{Seff1}
\end{equation}
The continuum representation (\ref{Seff}) determines the way in which the
original lattice model can be renormalized. Following to the standard
procedure \cite{Amit} we introduce the renormalization factors $Z_i^{\pm }$
for the ground-state parameters in the disordered and ordered phases by
relations 
\begin{eqnarray}
\pi _x &=&Z_x^{\pm }\pi _{xR},\;\pi _y=Z_y^{\pm }\pi _{yR}\;  \nonumber \\
t_{\pm } &=&(Z_2^{\pm }/Z)\,t_{\pm R},\;g=(Z_4^{\pm }/Z^2)g_R  \label{Ren}
\end{eqnarray}
where the indices $R$ denote quantum-renormalized quantities, 
\begin{equation}
g=K_{4-\varepsilon }L_\varepsilon \mu ^{-\varepsilon }u_{\text{cont}}
\end{equation}
is the coupling constant, $\mu $ is a parameter with the dimensionality of
inverse length,$\;K_d=[2^{d-1}\pi ^{d/2}\Gamma (d/2)]^{-1}$, and the factor $%
L_\varepsilon =\Gamma (1+\varepsilon /2)\Gamma (1-\varepsilon /2)$ [$\Gamma
(z)$ is the Euler gamma function] ensures the applicability of the one-loop
order results for not small $\varepsilon $ \cite{MSS}. For further treatment
it is useful to represent the renormalization factors as 
\begin{equation}
Z_i^{\pm }=Z_{Li}^{\pm }(g)Z_i^{\text{cont}}(g_R,\mu )  \label{ZZ}
\end{equation}
where $Z_i^{\text{cont}}$ are the corresponding factors for the continuum
model (\ref{Seff}) that contain divergent terms (which are independent of
lattice structure etc.) and $Z_{Li}$ all the others (lattice dependent)
corrections. It is important that the factors $Z_{Li}$ do not contain
divergences.

The expressions for $Z$-factors in the continuum model (\ref{Seff}) are well
known (see, e.g., Ref.\cite{Amit}). We use the cutoff scheme with cutting
integrals over quasimomentum at $\Lambda .$ Then to one-loop order we have 
\begin{eqnarray}
Z_x^{\text{cont}} &=&Z_y^{\text{cont}}=1+{\cal O}(g_R^2),\,\,  \nonumber \\
Z_2^{\text{cont}} &=&1+\frac{g_R}{2\varepsilon }\left( 1-\frac{\mu
^\varepsilon }{\Lambda ^\varepsilon }\right) ,\,  \nonumber \\
\,Z_4^{\text{cont}} &=&1+\frac{3g_R}{2\varepsilon }\left( 1-\frac{\mu
^\varepsilon }{\Lambda ^\varepsilon }\right) .  \label{Zi}
\end{eqnarray}
For our purposes it is convenient to put $\Lambda =(2d)^{1/2}$ (the lattice
constant is assumed to be equal to unity),\ rather than to pass to the limit 
$\Lambda \rightarrow \infty $ (as it is usual in the quantum field theory).
The expressions for $Z_{Li}$ can be deduced by comparison the above results
of perturbation theory for the original lattice model (Sections \ref{Sec2}
and \ref{Sec3}) and standard perturbation results for the continuum model (%
\ref{Seff}), see Ref. \cite{Amit}. We obtain 
\begin{eqnarray}
Z_{Lx} &=&Z_{Ly}=1\;  \nonumber \\
(Z_{L2}^{\pm })^{-1} &=&1+A_{\pm }(t_{\pm })+\frac g{2\varepsilon }\left( 
\frac 1{t_{\pm }^{\varepsilon /2}}-1\right) ,  \nonumber \\
\;(Z_{L4}^{-})^{-1} &=&1+A_{-}(t_{-})-B(t_{-})+\frac{3g}{2\varepsilon }%
\left( \frac 1{t_{-}^{\varepsilon /2}}-1\right)  \label{ZLi}
\end{eqnarray}
Note that, unlike the factors $Z_i^{\text{cont}}$, the quantities (\ref{ZLi}%
) are defined only for integer $\varepsilon .$ As follows from (\ref{ZZ}),
the determination of factors $Z_{Li}$ enables one to consider the continuum
model (\ref{Seff}) with the parameters 
\begin{eqnarray}
m^2 &=&(Z_{L2}^{\pm })^{-1}m_{\text{cont}}^2,\;u=(Z_{L4}^{\pm })^{-1}u_{%
\text{cont}}  \nonumber \\
c &=&c_0\left( 1+1/2S-X_0\right)
\end{eqnarray}
instead of the original lattice one. Thus the factors $Z_{Li}$ represent the
corrections owing to passing from the cutoff scheme in the original lattice
model to that in the continuum model, cf. Ref. \cite{Chakraverty}.

The flow functions for the coupling constant and energy gap have the
standard form \cite{Amit} 
\begin{eqnarray}
\beta (g_R) &=&\mu \frac{\partial g_R}{\partial \mu }=-\varepsilon g_R+\frac 
32g_R^2  \nonumber \\
\gamma (g_R) &=&\mu \frac{\partial \ln Z_2^{\text{cont}}}{\partial \mu }=-%
\frac 12g_R
\end{eqnarray}
The effective-Hamiltonian parameters $g_\rho ,\,t_\rho $ at the scale $\mu
^{\prime }=\mu \rho $ as determined by these flow functions read 
\begin{eqnarray}
g_\rho &=&\left[ 1+\frac{g_R}{g^{*}}(\rho ^{-\varepsilon }-1)\right]
^{-1}\rho ^{-\varepsilon }g_R\,\,  \nonumber \\
t_\rho &=&\left[ 1+\frac{g_R}{g^{*}}(\rho ^{-\varepsilon }-1)\right]
^{-1/3}t_{\pm R}  \label{Scal}
\end{eqnarray}
where $g^{*}=2\varepsilon /3$ is the stable fixed point to one-loop order.

We start the scaling procedure at $\mu =\Lambda $ and stop it at $\mu
^{\prime }=\Lambda t_{\pm }^{1/2}$ (thus $\rho =t_{\pm }^{1/2}$). For $%
\Delta _{\pm }$ and $\overline{S}$ we obtain the results 
\begin{eqnarray}
\Delta _{\pm R}^2(t_{\pm },0) &=&\frac 1{Z_{L2}^{\pm }}\frac{t_{\pm }}{%
\left[ 1+(3g_R/2\varepsilon )(1/t_{\pm }^{\varepsilon /2}-1)\right] ^{1/3}}
\label{dmrg} \\
\overline{S}_R(t_{-},0) &=&t_{-}^{1/2}\sqrt{\frac{Z_{L4}^{-}}{Z_{L2}^{-}}}%
\left[ 1+\frac{3g_R}{2\varepsilon }\left( \frac 1{t_{-}^{\varepsilon /2}}%
-1\right) \right] ^{1/3}
\end{eqnarray}
where, according to (\ref{Ren}), (\ref{Zi}), 
\begin{equation}
g_R=(Z_{4L}^{\pm })^{-1}(2d)^{-\varepsilon /2}K_{4-\varepsilon }L_\varepsilon
\end{equation}
In the 3D case 
\begin{equation}
\frac 1\varepsilon \left( \frac 1{t^{\varepsilon /2}}-1\right) \rightarrow 
\frac 12\ln \frac 1t
\end{equation}
so that the QPT critical exponents for the order parameter and the gap
(inverse correlation length) are Gaussian one, 
\begin{equation}
\beta =1/2,\,\,\,\nu =1/2,  \label{CE3}
\end{equation}
and logarithmic corrections are present. At the same time, in the 2D case we
obtain 
\begin{equation}
\beta =1/3,\,\,\,\nu =7/12  \label{CE2}
\end{equation}
which are standard one-loop results for the one-component $\phi ^4$ theory
in $d+1=3$ dimensions.

In a strong enough longitudinal magnetic field ($1-\Omega /\Omega _c\ll
\theta _H^2\ll 1$) we obtain 
\begin{eqnarray}
\Delta _{-R}^2(\theta _H,0) &=&\frac 2{3Z_{L2}^{\prime }}\frac 1{\left[
1+(3g_R/2\varepsilon )(1/\theta _H^\varepsilon -1)\right] ^{1/3}}  \nonumber
\\
\overline{S}_R(\theta _H,0) &=&\theta _H\sqrt{\frac{Z_{L4}^{^{\prime }}}{%
Z_{L2}^{\prime }}}\left[ 1+\frac{3g_R}{2\varepsilon }\left( \frac 1{\theta
_H^\varepsilon }-1\right) \right] ^{1/3}
\end{eqnarray}
where 
\begin{eqnarray}
Z_{L2}^{^{\prime }} &=&1+A^{\prime }(\theta _H^2)+\frac g{2\varepsilon }%
\left( \frac 1{\theta _H^\varepsilon }-1\right) ,  \nonumber \\
\;Z_{L4}^{^{\prime }} &=&1+A^{\prime }(\theta _H^2)-B^{\prime }(\theta _H^2)+%
\frac{3g}{2\varepsilon }\left( \frac 1{\theta _H^\varepsilon }-1\right)
\end{eqnarray}
Thus, as well as for the dependences of ground-state properties on $t,$ in
the 3D case one has the mean-field value 
\begin{equation}
\delta =3
\end{equation}
and the logarithmic corrections are present. At $d=2$ we obtain the critical
exponent 
\begin{equation}
\delta =9/2
\end{equation}
Note that the scaling relations at $d=2$ are slightly violated since the
corresponding value of $\varepsilon $ is in fact not small and the $%
\varepsilon $-expansion is applicable with a poor accuracy. However, this
violation is not too large (the value $\delta =5$ can be calculated taking
into account that the critical exponent $\eta =0$ to one-loop order), which
indicates that the one-loop approximation gives adequate results even in
this case.

The ground-state parameters at zero longitudinal magnetic field are shown
and compared with RPA and GSPT results in Fig.1 for the 3D case and in Fig.
2 for the 2D case. The corrected values of $\Omega _c$ obtained from GSPT
(see above), instead of those from first-order $1/S$ expansion, are used in
the calculations. The $1/S$-results for $\Omega _c$ are marked by arrows.
One can see that, unlike results of RPA and $1/S$-expansions, RG results
have a correct critical behavior with critical exponents given by (\ref{CE3}%
) and (\ref{CE2}); besides that, they are very close to GSPT result for $%
d=3. $ For $d=2$ the difference between RG$^{\prime }$ and GSPT results
increases, which demonstrates that the $\varepsilon $-expansion has a poor
accuracy here. Far from the quantum phase transition, the RG results
coincide with those of $1/S$ perturbation theory.

\subsection{\label{Sect.4}Finite-temperature properties near QPT}

At finite temperature the situation is more complicated, since not only
``spin-wave'' excitations, considered in previous sections, contribute to
thermodynamic properties. At $\Omega /\Omega _c\ll 1$ we have $T_c\sim IS^2$
and the phase transition occurs due to vanishing of $\langle |{\bbox \pi }%
|\rangle .$ The dominant excitations in this case are domain walls. Another
situation occurs near QPT ($1-\Omega /\Omega _c\ll 1$) where the temperature
phase transition is connected with the rotation of $\langle {\bbox \pi }%
\rangle $ in the spin space, while its absolute value is only slightly
changed with temperature. The dominant excitations here are the
``spin-wave'' excitations, except for a narrow critical region close to $%
T_c. $ At intermediate values of $\Omega \ $both effects, the rotation of $%
\langle {\bbox \pi }\rangle $ and temperature variation of its absolute
value, are important. Thus the $1/S$-expansion can be applied to describe
the temperature phase transition only near QPT.

Being rewritten through the quantum-renormalized ground-state parameters,
finite-temperature properties near QPT are universal. The finite-temperature
order parameter and energy gaps obey the scaling laws 
\begin{mathletters}
\begin{eqnarray}
\overline{S}^2(t_{-},T) &=&\overline{S}^2(t_{-},0)\,f\left( \frac T{c\Delta
_{-0}}\right)  \label{Scal1} \\
\Delta _{\pm }(t_{\pm },T) &=&\Delta _{\pm }(t_{\pm },0)g_{\pm }\left( \frac 
T{c\Delta _{\pm 0}}\right)  \label{Scal2}
\end{eqnarray}
where $\Delta _{\pm 0}=\Delta _{\pm }(t,0),$ $f$ and $g_{\pm }$ are
universal scaling functions with $f(0)=g_{\pm }(0)=1$. The transition
temperature is determined by zero of the function $f\left( T/c\Delta
_{-0}\right) $ or, equivalently, of $g_{-}\left( T/c\Delta _{-0}\right) .$
As discussed in Ref.\cite{Sachdev}, the functions $g_{+}(x)$ and $g_{-}(x)$
are connected by the procedure of analytical continuation. Due to
universality of scaling functions (\ref{Scal1}), (\ref{Scal2}),\ the
continuum limit of developed theory, i.e. the action (\ref{Seff}), can be
used when treating the finite-temperature properties.

Now we pass to calculation of the functions $f$ and $g_{\pm }.$ Consider
first the perturbation approach. We obtain 
\end{mathletters}
\begin{equation}
\Delta _{+}^2(t_{+},T)=\Delta _{+}^2(t_{+},0)+\frac u{2d}\left( \frac{2T}c%
\right) ^{d-1}F_d\left( \frac{c^2d}{4T^2}t_{+}\right) 
\end{equation}
for the disordered phase and 
\begin{mathletters}
\begin{eqnarray}
\overline{S}_R^2(t_{-},T) &=&\overline{S}_R^2(t_{-},0)-\frac u{4d}\left( 
\frac{2T}c\right) ^{d-1}F_d\left( \frac{c^2d}{2T^2}t_{-}\right) 
\label{StPt} \\
\Delta _{-}^2(t_{-},T) &=&\overline{S}_R^2(t_{-},T)\left[ \frac{\Delta
_{-}^2(t_{-},0)}{\overline{S}_R^2(t_{-},0)}+\frac{3u}4\left( \frac{2T}c%
\right) ^{d-3}F_d^{\prime }\left( \frac{c^2d}{2T^2}t_{-}\right) \right] 
\label{DmPt}
\end{eqnarray}
for the ordered phase, where 
\end{mathletters}
\[
F_d(x)=K_d\int\limits_0^\infty \frac{q^{d-1}dq}{\sqrt{q^2+x}}\left( \coth 
\sqrt{q^2+x}-1\right) ,\,\,\,F_3(0)=\frac 1{24}
\]
and $F_d^{\prime }(x)$ is the derivative with respect to $x$. Thus we have
for the static susceptibility in the disordered phase at $I\Delta _{+}\ll
T\ll I$ \cite{Millis,Sachdev} 
\begin{equation}
\chi ^{xx}=\frac 1{I_0}\frac 1{\Delta _{+}^2(t,0)+\gamma (2T/c)^{d-1}}
\label{HiDis}
\end{equation}
with $\gamma =3I_0F_d\left( 0\right) /4c.$

At $T\geq I$ thermodynamic properties cannot be determined correctly from
the above approach since in this temperature region higher-order terms in
the $1/S$-expansion contribute to partition function and such an expansion
becomes inapplicable. However, one can expect that at $T\gg I$ the
thermodynamics is the same as for the well-studied Ising model. In
particular, the susceptibility obeys the Curie law 
\begin{equation}
\chi ^{xx}=\frac{S(S+1)}{3T}  \label{CurieLaw}
\end{equation}
on both sides of QPT.

The equation for the transition temperature reads 
\begin{equation}
\overline{S}^2(t_{-},0)=\frac u{4d}\left( \frac{2T_c}c\right)
^{d-1}F_d\left( \frac{c^2d}{2T_c^2}t_{-}\right)  \label{EqTc}
\end{equation}
Since at small $t_{-}$ and $d\leq 3$ one has from (\ref{Sl})$\,\overline{S}%
^2(t_{-},0)\propto (t_{-}{})^{(d-1)/2},$ we obtain 
\begin{equation}
T_c\propto \sqrt{t_{-}},\,\,\,\,1<d\leq 3  \label{TcSq}
\end{equation}
where the coefficient of proportionality is determined by the solution of
Eq. (\ref{EqTc}). For $d>3$ we derive 
\begin{equation}
T_c\propto (t_{-})^{1/(d-1)},\,\,d>3
\end{equation}
The mean-field logarithmic behavior (\ref{TcMF}) is reproduced only for $%
d\rightarrow \infty $.

Consider now the renormalization of the finite-temperature properties at $%
d\leq 3$. To this end we use the approach of Ref. \cite{Sachdev}, which
treats the renormalization of the effective classical model. The disordered
phase was considered in details in Ref. \cite{Sachdev}. Instead of
analytical continuation of these results to ordered phase, we perform direct
calculation of finite-temperature properties in the ordered phase. This
gives a possibility to calculate correctly the value of $T_c$ not too close
to QPT and also to describe finite-temperature properties at $T<T_c$. The
generalization of the approach of Ref. \cite{Sachdev} to the ordered phase
is trivial. We integrate out all the modes with nonzero Matsubara
frequencies from finite-temperature partition function to obtain the
effective action for the field 
\begin{equation}
\Pi =\int_0^{c/T}d\tau \,\pi _x,
\end{equation}
which corresponds to $\omega _n=0$ mode, in the form 
\begin{eqnarray}
{\cal S}_{\text{cl}} &=&\frac c{2T}\int d^dr\left[ K(\nabla \widetilde{\Pi }%
)^2+R\widetilde{\Pi }^2\right] +\frac{\overline{\Pi }}{3!}\frac{cU}T\int d^dr%
\widetilde{\Pi }^3  \nonumber \\
&&\ \ \ \ \ \ \ \ +\frac 1{4!}\frac{cU}T\int d^dr\widetilde{\Pi }^4+...
\label{Scl1}
\end{eqnarray}
Here $\widetilde{\Pi }=\Pi -\overline{\Pi }$, $\overline{\Pi }$ is
determined by the condition of absence in (\ref{Scl1}) of terms that are
linear in $\widetilde{\Pi }$. The parameters of the model (\ref{Scl1}) are
given by 
\begin{equation}
R(T)=\frac 13U(T)\,\overline{\Pi }^2(T)
\end{equation}
and for $d=3$%
\begin{eqnarray}
\overline{\Pi }^2(T) &=&\frac{18}u\left[ \overline{S}_R^2(t_{-},0)-\frac 13%
\frac{8\pi ^2g_R}{1+(3g_R/2)\ln (1/\Delta _{-0})}\left( \frac Tc\right) ^2%
\widetilde{F}_3\left( \frac{3\Delta _{-0}^2c^2}{2T^2}\right) \right] ,
\label{RR} \\
U(T) &=&\frac{8\pi ^2g_R}{1+(3g_R/2)\ln (1/\Delta _{-0})}\left[ 1+\frac{6\pi
^2g_R}{1+(3g_R/2)\ln (1/\Delta _0)}\widetilde{F}_3^{\prime }\left( \frac{%
3\Delta _{-0}^2c^2}{2T^2}\right) \right] ,  \label{RU}
\end{eqnarray}
where 
\begin{equation}
\widetilde{F}_3(x)=K_3\int\limits_0^\infty q^2dq\left[ \frac{\coth \sqrt{%
q^2+x}-1}{\sqrt{q^2+x}}-\frac 1{q^2+x}+\frac 1{q^2}\right] ,
\end{equation}
$\widetilde{F}_3^{\prime }(x)$ means the derivative with respect to $x,$ and
we have represented (\ref{RR}) and (\ref{RU}) in the scaling form by
replacing $t_{-}\rightarrow \Delta _{-0}^2$ in arguments of $F_3(x),$ $%
F_3^{\prime }(x)$. Near QPT (i.e. for small $\Delta _{-0}$) function $R(T)\ $%
coincides with that determined by continuation from paramagnetic phase, as
it should be. The value of $K(T)$ will be needed only in zeroth-loop order, $%
K=1.$

The critical temperature is determined by the condition $\overline{\Pi }%
(T_c)=0$. Closely enough to the critical point (at $\ln (1/\Delta _{-0})\gg
1 $) we have 
\begin{equation}
T_c=\frac 3{2\pi }c\Delta _{-0}\sqrt{6\ln (1/\Delta _{-0})}  \label{tcrg3d}
\end{equation}
in agreement with Ref. \cite{Sachdev} (our definition of $\Delta _{-}$
differs $(2d)^{1/2}c$ times from that used in Ref.\cite{Sachdev}). At the
same time, the expansion in the bare splitting (magnetic field) yields 
\begin{equation}
T_c\propto c\sqrt{t_{-}}\ln ^{1/3}(1/t_{-})\propto \overline{S}_R(t_{-},0)
\label{Tc1}
\end{equation}
where the coefficient of proportionality can be determined numerically from (%
\ref{dmrg}) and (\ref{tcrg3d}). Thus, due to ground-state renormalizations,
the dependences of $T_c$ on the bare and renormalized splittings turn out to
be different.

The resulting classical action (\ref{Scl}) is renormalized in a standard way 
\cite{Amit}. One can introduce the renormalization constants for
finite-temperature theory in the form 
\begin{equation}
R=(Z_2^T/Z^T)R_r,\,\,\Pi =Z^T\Pi _r,\,\,\,U=\frac{\mu ^\epsilon }{%
K_{4-\epsilon }L_\epsilon }(Z_4^T/Z^{T2})U_r
\end{equation}
where the index ``$r$'' stands for the quantities renormalized by
temperature fluctuations, and $\epsilon =1+\varepsilon $. The expressions
for $Z$-factors are the same as for the ground-state renormalization factors
(\ref{Zi}) with the replacement $\varepsilon \rightarrow \epsilon .$
Formulas of RG transformation also have the same form (\ref{Scal}) as for
the ground-state properties with $t\rightarrow R,$ $g\rightarrow U$ etc.
However, now already at $d=3$ ($\varepsilon =0$) we have $\epsilon =1$ and
thus the $\epsilon $-expansion can be used only approximately.

For the energy gap we obtain in this way the expression 
\begin{equation}
\Delta _{-}^2(t_{-},T)=\frac{R(T)}6\left[ 1+\frac{3TK_3L_1U(T)}{2cR^{1/2}(T)}%
\right] ^{-1/3}
\end{equation}
where we have put $\epsilon =1$. For the temperature-dependent magnetization
we obtain 
\begin{equation}
\overline{S}_R^2(t_{-},T)=\frac{uR(T)}{6U(T)}\left[ 1+\frac{3TK_3L_1U(T)}{%
2cR^{1/2}(T)}\right] ^{2/3}
\end{equation}
The values of the temperature-transition critical exponents, 
\begin{equation}
\beta _T=1/3,\,\,\,\nu _T=7/12,  \label{ExpT}
\end{equation}
coincide with those of 2D quantum phase transition (\ref{CE2}).

The calculated dependence $T_c(\Omega )$ is shown and compared with the
mean-field and HTSE\ results in Fig. 3. One can see that near QPT the
dependence $T_c(\Omega )$ calculated from (\ref{RR})\ is in excellent
agreement with HTSE data. At the same time, far from QPT our approach gives
much larger values of $T_c,$ as discussed in the beginning of the present
section. The inflection point of the curve $T_c(\Omega ),\,\Omega
^{*}=0.35I_0,$ may be approximately related to the transverse-field value
where the ``non-spin-wave'' excitations becomes important for description of
finite-temperature properties.

For $d=2$ the system is far from its upper critical dimensionality ($%
\epsilon =2$) and $\epsilon $-expansion becomes inapplicable. Therefore we
can perform only ground-state renormalizations in the results of
perturbation theory (\ref{StPt}) and (\ref{DmPt}). In this case the critical
exponents of the temperature phase transition still have their Gaussian
values. However, universality hypothesis predicts that the temperature phase
transition critical exponents coincide with those for the 2D Ising model, 
\begin{equation}
\beta _T=1/8,\;\nu _T=1.
\end{equation}
With account of the ground-state renormalizations, the result (\ref{TcSq})
for the critical temperature near the quantum phase transition ($1/\Delta
_{-0}\gg 1$) takes the form 
\begin{equation}
T_c\propto \Delta _{-0}  \label{Tc2D-1}
\end{equation}
while 
\begin{equation}
T_c\propto (t_{-})^{5/12}  \label{Tc2D-2}
\end{equation}
in terms of bare splitting (or external transverse magnetic field). The
correct description of termodynamics below $T_c$ in the 2D case is still an
open problem.

\section{The Heisenberg model with strong easy-plane anisotropy}

\subsection{Ground-state properties}

We start from the general Hamiltonian of a spin system in crystal field
which induces the single-site anisotropy, 
\begin{equation}
{\cal H}=V_{\text{cf}}-\frac{{\cal I}}2\sum_{\langle ij\rangle }{\bf J}_i%
{\bf J}_j  \label{Hcf}
\end{equation}
where $V_{\text{cf}}$ is the crystal field potential, ${\bf J}$ are momentum
operators, ${\cal I}$ is the exchange integral, and the direction of spin
alignment will be supposed along $z$-axis. In this Section we consider the
single-site easy-plane anisotropy which corresponds to 
\begin{equation}
V_{\text{cf}}=D\sum_i(J_i^x)^2  \label{E-pl}
\end{equation}
where $D>0$ is the anisotropy parameter. For integer values of $J$ the
lowest level is singlet. In this case with increasing $D$ the model (\ref
{E-pl})\ demonstrates at some value $D_c$ a second-order phase transition
from the phase with collinear ferromagnetic order $\langle J^z\rangle \neq 0$
to disordered phase. At the same time, the quadrupole order parameter 
\begin{equation}
Q\equiv 3\langle (J^x)^2\rangle -J(J+1)
\end{equation}
is nonzero in both the phases. For half-integer values of $J,$ such
transition is absent since the lowest state is two-fold degenerate. In the
classical limit $J\rightarrow \infty $ with $J$ being integer, we have $D_{%
\text{c}}\sim J(J+1){\cal I}\rightarrow \infty ,$ so that integer and
half-integer values of $J$ become indistinguishable.

For integer $J$ the ground state is $|A\rangle =|\widetilde{0}\rangle ,\;$%
and first excited state is doublet $\,|B_1\rangle =|\widetilde{1}\rangle $, $%
|B_2\rangle =|-\widetilde{1}\rangle $ where $|\widetilde{M}\rangle $ are the
eigenstates of $J^x$. For $J=1$ passing to the eigenstates $|M\rangle $ of $%
J^z$ yields 
\begin{eqnarray}
|A\rangle &=&\frac 1{\sqrt{2}}(|1\rangle -|-1\rangle ),  \nonumber \\
\,\,|B_1\rangle &=&|0\rangle ,\,\,|B_2\rangle =\frac 1{\sqrt{2}}(|1\rangle
+|-1\rangle )
\end{eqnarray}

To consider the vicinity of QPT, we have to generalize the theory developed
in previous Section on the singlet-doublet case. Further we restrict
ourselves to the case $J=1$. In the initial spin space, QPT in the model (%
\ref{Hcf}) with (\ref{E-pl}) is not of orientational character: the spins
always lie in the easy-plane. Thus the spin-wave theory in its standard form
cannot properly describe the model (\ref{Hcf}) near such a transition (see
e.g. Ref. \cite{Chub}). However, as discussed in Refs. \cite{Onufr,Valkov},
this transition can be viewed as orientational one in the complete $SU(3)$
space which includes the $SU(2)$ spin subspace. Most convenient way to
consider the rotations in the extended $SU(3)$ space is to rewrite the
Hamiltonian (\ref{Hcf}) with the crystal field (\ref{E-pl}) in terms of the
Hubbard operators $X_i^{mn}=|m_i\rangle \langle n_i|$, 
\begin{eqnarray}
{\cal H} &=&\frac D2\sum_i(X_i^{00}+X_i^{1,-1}+X_i^{-1,1})  \nonumber \\
&&\ \ \ \ \ \ \ \ \ \ \ \ \ \ \ -\frac{{\cal I}}2\sum_{\langle ij\rangle
}\left[
(X_i^{10}+X_i^{0-1})(X_j^{01}+X_j^{-10})+(X_i^{11}-X_i^{-1,-1})(X_i^{11}-X_j^{-1,-1})\right]
\label{He-pl}
\end{eqnarray}
The rotation through ``angle'' $\theta $ in $SU(3)$ space is performed by
the unitary transformation operator $U(\theta )$ (see Appendix C).

Following to strategy described in Sect.2, we define the ground-state
critical value of $D_c$ from the condition $\sin \theta =1$ which yields 
\begin{equation}
\frac{D_c}{2{\cal I}_0}=1+\frac{3\lambda }2-\lambda \sum_{{\bf k}}\frac{6%
{\cal I}_0+{\cal I}_{{\bf k}}+{\cal I}_{{\bf k}}^2/{\cal I}_0}{2E_{{\bf k}}^0%
}  \label{Dcr}
\end{equation}
where $E_{{\bf k}}^0=2\sqrt{{\cal I}_0({\cal I}_0-{\cal I}_{{\bf k}})}%
,\;\lambda (=1)$ is the formal expansion parameter$.$ The critical value
obtained from (\ref{Dcr})\ in the 3D case is $D_c/2{\cal I}_0=0.73$, which
coincides with the result of HTSE \cite{HTSE-Pl}. For the ground-state
magnetization we obtain 
\begin{eqnarray}
\langle J^z\rangle _{T=0}^2 &=&t_{-}\left[ 1+B(t_{-})\right] \\
B(t_{-}) &=&-\frac \lambda {t_{-}}\sum_{{\bf k}}\left[ 2\frac{2{\cal I}%
_0+(2-\eta ^2){\cal I}_{{\bf k}}}{E_{{\bf k}\alpha }}\right.  \nonumber \\
&&\ \ \ \ \left. +\frac{(1+\eta ){\cal I}_0-{\cal I}_{{\bf k}}+\eta {\cal I}%
_{{\bf k}}^2/{\cal I}_0}{E_{{\bf k}\beta }}-\frac{6{\cal I}_0+{\cal I}_{{\bf %
k}}+{\cal I}_{{\bf k}}^2/{\cal I}_0}{E_{{\bf k}}^0}\right]  \label{SD}
\end{eqnarray}
where the excitation spectrum $E_{{\bf k}\alpha ,\beta }$ is given by (\ref
{E-plEka}), (\ref{E-plEkb}), and $t_{-}=1-(D/D_c)^2,$ $\eta =(1-t_{-})^{1/2}$
The excitations of $\alpha $-type have a gap; they are analogous to the
excitations in the transverse-field Ising model, considered in previous
section. The excitations of $\beta $-type are gapless due to spontaneous
breaking of rotational symmetry in the $y$-$z$ plane of spin space; these
excitations are specific for $n\geq 2$ systems. Near QPT we have $E_{{\bf k}%
\beta }\approx E_{{\bf k}}^0$ and we return to the perturbation result (\ref
{Magn})\ for the transverse-field Ising model with $E_{{\bf k}}=E_{{\bf k}%
\alpha }$ being the critical mode. However, the renormalization of (\ref{SD}%
) is performed in a different way in comparison with the transverse-field
Ising model because of another symmetry of the model (see below). The energy
gap $\Delta _{-}$ in the ordered phase is given by {\ 
\begin{eqnarray}
\Delta _{-}^2(t_{-},0) &=&t_{-}\left[ 1+A(t_{-})\right] , \\
A(t_{-}) &=&\frac 1{t_{-}}(A_0+A_1)-1  \label{DD}
\end{eqnarray}
where }$A_{0,1}$ are determined by (\ref{A0}) and (\ref{A1}).

In the presence of longitudinal magnetic field, i.e. of the field $H$,
directed along $z$-axis, both modes $\alpha $ and $\beta $ becomes gapped,
since this field breaks the rotational symmetry. As well as for the
transverse-field Ising model, we can expect that at the intermediate
magnetic field values$\,1-D/D_c\ll (H/D_c)^{2/3}\ll 1$ the ground-state
properties near QPT will be determined by the magnetic field rather than by $%
t_{\pm }$. In this region we obtain the energy spectra 
\begin{eqnarray}
E_{{\bf k}\alpha }^2 &=&4{\cal I}_0\left[ {\cal I}_0(1+\theta _H^2)-{\cal I}%
_{{\bf k}}(1-\theta _H^2/2)\right]  \nonumber \\
E_{{\bf k}\beta }^2 &=&\left[ 2({\cal I}_0-{\cal I}_{{\bf k}})+\theta _H^2(%
{\cal I}_0+{\cal I}_{{\bf k}})/2\right] \left[ 2{\cal I}_0+\theta _H^2({\cal %
I}_0-{\cal I}_{{\bf k}})/2\right]
\end{eqnarray}
where $\theta _H=(4H/D_c)^{1/3}$. Performing the calculations which are
similar to those for the transverse-field Ising model, we obtain the result 
\begin{equation}
\overline{S}=\theta _H\left[ 1+B^{\prime }(\theta _H^2)\right]
\end{equation}
where 
\begin{eqnarray}
B^{\prime }(\theta _H^2) &=&-\frac \lambda {3\theta _H^2}\sum_{{\bf k}%
}\left[ 2\frac{2({\cal I}_0+{\cal I}_{{\bf k}})(1+\theta _H^2/2)-{\cal I}_{%
{\bf k}}(1-\theta _H^2)}{E_{{\bf k}\alpha }}\right.  \nonumber \\
&&+\frac{(2+\theta _H^2/2){\cal I}_0-{\cal I}_{{\bf k}}+(1-\theta _H^2/2)%
{\cal I}_{{\bf k}}^2/{\cal I}_0}{E_{{\bf k}\beta }}\   \nonumber \\
&&\left. -\frac{6{\cal I}_0+{\cal I}_{{\bf k}}+{\cal I}_{{\bf k}}^2/{\cal I}%
_0}{E_{{\bf k}}^0}\right]
\end{eqnarray}

A more complicated situation takes place in the case of the transverse field
directed along $x$-axis \cite{Onufr,Chub}. This field induces a deviation of
spins from easy plane. With increasing the field value there occurs a
cascade of $J$ second-order phase transitions from ferromagnetically ordered
phases with $\langle J^z\rangle \neq 0,$ $\langle J^x\rangle \neq 0$ to
phases which are ordered only along $x$-axis ($\langle J^z\rangle =0,$ $%
\langle J^x\rangle \neq 0$) and vice versa. The reason for this is the
modification of level scheme in the magnetic field directed along the hard
axis: in the case where lowest state is doublet the long-range order along $%
z $-axis is present, while in the case of singlet ground state it is
evidently absent. We do not consider these transitions here (see discussion
of such transition in Refs. \cite{Onufr,Chub,HTr}).

\subsection{Ground-state renormalizations}

The above theory can be easily reformulated in the path integral formalism.
The partition function has the form 
\begin{equation}
{\cal Z}=\int D[a,a^{\dagger },b,b^{\dagger }]\exp \left\{ a^{\dagger }\frac{%
\partial a}{\partial \tau }+b^{\dagger }\frac{\partial b}{\partial \tau }-%
{\cal H}(a,a^{\dagger },b,b^{\dagger })\right\}
\end{equation}
where ${\cal H}(a,a^{\dagger },b,b^{\dagger })$ is the average of the boson
Hamiltonian over the coherent states $|a,b\rangle $ \cite{Klauder} (see also
Ref. \cite{ArovasBook}). The continuum limit of the theory can be obtained
if we introduce real variables $\pi _{x,y}$ and $Q_{x,y}$ instead of the
complex ones $a,b$ by the relations 
\begin{eqnarray}
a &=&\pi _x+iQ_x  \nonumber \\
b &=&Q_y+i\pi _y
\end{eqnarray}
(Note that $\pi _x$ and $\pi _y$ correspond to $S^z$ and $S^y$ in the
original spin space, and two additional variables $Q_{x,y}$ arise due to
passing from $SU(2)$ to $SU(3)$ space). We obtain 
\begin{eqnarray}
{\cal S}_{\text{cont}} &=&\frac 12\int\limits_0^{c/T}d\tau \int d^dr\left[
-2i\widetilde{Q}_x(\partial \widetilde{\pi }_x/\partial \tau )+2i\widetilde{Q%
}_y(\partial \widetilde{\pi }_y/\partial \tau )\right.  \nonumber \\
&&\ \ \ \ \ \ \ \left. +\widetilde{{\bf Q}}^2+(\nabla \widetilde{\bbox{\pi }}%
)^2+m^2\widetilde{\bbox{\pi }}^2\right] +\frac u{4!}\int\limits_0^{c/T}d\tau
\int d^dr\widetilde{\bbox{\pi }}^4  \nonumber \\
&&\ \ +h\int\limits_0^{c/T}d\tau \int d^dr\,\widetilde{\pi }_x  \label{Seff2}
\end{eqnarray}
where we have introduced the notations $\widetilde{\bbox{\pi }}^2=({\cal I}%
/c_0)\bbox{\pi }^2,\;\widetilde{{\bf Q}}^2=({\cal I}_0/c_0){\bf Q}^2$, $h=H/(%
{\cal I}c_0)^{1/2},$ the bare spin-wave velocity is given by $c_0=2\sqrt{2d}%
{\cal I}$ and we have included into (\ref{Seff2}) the term connected with
the external magnetic field $H$ along the $S^z$ axis. The parameters of this
model, determined by the continuum limit, read 
\begin{eqnarray}
m_{\text{cont}}^2 &=&-t_{-}d  \nonumber \\
u_{\text{cont}} &=&6d(c_0/{\cal I)}\lambda \zeta  \nonumber \\
c_{\text{cont}} &=&c_0
\end{eqnarray}
with $\zeta =1-t_{-}$ in the ordered phase under consideration. Proceeding
in the same way as in the previous Section we integrate over $Q_{x,y}$. Then
we obtain the action of the standard two-component quantum $\phi ^4$-theory
in an external field, 
\begin{eqnarray}
{\cal S}_{\text{cont}} &=&\frac 12\int\limits_0^{c/T}d\tau \int d^dr\left[
\,(\partial \widetilde{\bbox{\pi }})^2+m^2\widetilde{\bbox{\pi }}^2\right] 
\nonumber \\
&&\ \ \ \ +\frac u{4!}\int\limits_0^{c/T}d\tau \int d^dr\widetilde{\bbox{\pi
}}^4+h\int\limits_0^{c/T}d\tau \int d^dr\,\widetilde{\pi }_x.  \label{S2}
\end{eqnarray}

There is a crucial difference from the one-component model of the previous
section, which is due to existence in the ordered phase of the gapless
Goldstone mode at $H=0$. This mode changes the renormalization conditions
since it leads to infrared divergences\cite{GIR}. To treat these
divergences, we take the value of magnetic field $H$ finite, but small
enough to satisfy $(H/D_c)^{2/3}\ll 1-D/D_c$. The renormalization of the
action (\ref{S2}) is considered in Appendix D. We obtain for effective
Hamiltonian parameters at the scale $\Lambda \rho ,$ $\Lambda =(2d)^{1/2}$
the results ($d=3$) 
\begin{eqnarray}
g_\rho ^{-1} &=&g_R^{-1}\left[ 1+(3g_R/4)\ln (1/t_{-})+(g_R/6)\ln (1/\rho
)\right]  \nonumber \\
t_\rho ^{-1} &=&t_R^{-1}\left[ 1+(3g_R/4)\ln (1/t_{-})+(g_R/6)\ln (1/\rho
)\right]  \nonumber \\
&&\ \ \ \ \ \ \ \ \times \left[ 1+(5g_R/6)\ln (1/t_{-})\right] ^{-3/5}\Phi
_0(g_R,t_{-}^2)  \label{scal2}
\end{eqnarray}
where the function $\Phi _0(g,x)$ is given by (\ref{F0}), 
\begin{equation}
g_R=K_4^{-1}Z_{L4}^{-}u_{\text{cont}}
\end{equation}
is the renormalized coupling constant. For the non-universal $Z$-factors we
have 
\begin{eqnarray}
Z_L &=&1\;  \nonumber \\
(Z_{L2}^{-})^{-1} &=&1+\widetilde{A}(t_{-})+\frac g4\ln \frac 1{t_{-}}, 
\nonumber \\
\;(Z_{L4}^{-})^{-1} &=&1+\widetilde{A}(t_{-})-B(t_{-})+\frac{3g}4\ln \frac 1{%
t_{-}},
\end{eqnarray}
where $A(t_{-}),$ $B(t_{-})$ are given by (\ref{SD}), (\ref{DD}), the tilde
sign means that the contributions of the Goldstone mode $\beta $ should be
excluded from $A(t_{-})$.

Putting in the above expressions $\rho =\widetilde{h}^{1/2}$ where $%
\widetilde{h}=H/[{\cal I}_0\overline{J}_R(H=0)]$ we have for the
magnetization at $d=3$ the result 
\begin{equation}
\overline{J}_R(t_{-},0)=\sqrt{\frac{Z_{L4}^{-}t_{-}}{Z_{L2}^{-}}}\left[ 1+%
\frac{5g_R}6\ln \frac 1{t_{-}}\right] ^{3/10}\Phi _0^{1/2}(g_R,t_{-}^2)
\end{equation}
(the terms divergent in $H$ are canceled in $\overline{J}_R$). The gap for $%
\alpha $-type excitations, which determines the longitudinal susceptibility,
reads 
\begin{mathletters}
\begin{eqnarray}
\Delta _{-}^2(t_{-},0) &=&12\Theta _0^2/\ln (1/\widetilde{h})  \label{dmrg3d}
\\
\Theta _0^2 &=&\frac{t_{-}}{Z_{L2}^{-}}\left[ 1+\frac{5g_R}6\ln \frac 1{t_{-}%
}\right] ^{3/5}\frac{\Phi _0(g_R,t_{-}^2)}{g_R}  \nonumber \\
\ &=&Z_{L4}^{-}\frac{\overline{J}_R^2(t_{-},0)}{g_R}  \nonumber
\end{eqnarray}
Up to some nonuniversal factor $Z_\rho $ we have in the one-loop order $%
\Theta _0=Z_\rho (\rho _s/6dc)^{1/2}$ with $\rho _s$ being the ground-state
spin stiffness. At $H\rightarrow 0$ the gap vanishes as $\ln ^{-1}({\cal I}%
_0/H)$ $,$ which is a consequence of degeneracy of the system.

For intermediate values of external magnetic field, i.e. at $1-D/D_c\ll
(H/D_c)^{2/3}\ll 1$, a characteristic scale for both types of excitations is 
$1/\theta _H$, and the expressions for renormalization factors $Z_i^{\text{%
cont}}$ have the form, which is standard in the two-component $\phi ^4$ model%
\cite{Amit}. Then we obtain for the magnetization 
\end{mathletters}
\begin{equation}
\overline{J}_R(\theta _H,0)=\theta _H\sqrt{\frac{Z_{L4}^{\prime }}{%
Z_{L2}^{\prime }}}\left[ 1+(5g_R/6)\ln (1/\theta _H^2)\right] ^{3/10}
\end{equation}
with 
\begin{equation}
Z_{L4}^{^{\prime }}/Z_{L2}^{^{\prime }}=1-B^{\prime }(\theta _H^2)+g\ln 
\frac 1{\theta _H},
\end{equation}

\subsection{Finite-temperature properties}

Using perturbation theory we obtain for the finite-temperature magnetization
the result (see Appendix C) 
\begin{equation}
\langle J^z\rangle ^2=\langle J^z\rangle _{T=0}^2-\frac{{\cal I}_0\lambda }{%
2c_0}\left( \frac{2T}{c_0}\right) ^{d-1}\left[ 3F_d\left( \frac{c_0^2d}{2T^2}%
t_{-}\right) +F_d(0)\right]  \label{JzT}
\end{equation}
where $t_{-}=1-(D/D_c)^2.$ The first and second terms in the square brackets
correspond to contributions of $\alpha $ and $\beta $-type excitations
respectively. At extremely low temperatures $T\ll {\cal I}_0t_{-}$ the
contribution of $\alpha $-excitations is exponentially small and the
temperature-dependent part of magnetization is determined entirely by second
term in square brackets of (\ref{JzT}). At temperatures $T\gg {\cal I}%
_0t_{-} $ the situation changes and both type of excitations give the same
temperature dependence, the contribution of the mode $\alpha $ being three
times larger.

Consider now the renormalization of the finite-temperature theory.
Integrating out the field $\pi (q,\omega _n)$ with $\omega _n\neq 0$ from
the action (\ref{S2}) we obtain the action of effective classical model 
\begin{eqnarray}
{\cal S}_{\text{cl}} &=&\frac c{2T}\sum_{{\bf q}}\left\{ q^2\widetilde{{%
\bbox \Pi }}_{{\bf q}}\widetilde{{\bbox \Pi }}_{-{\bf q}}+\left[ R(T)+\frac{%
3h}{\overline{\Pi }(T)}\right] \widetilde{{\Pi }}_{x{\bf q}}\widetilde{{\Pi }%
}_{x,-{\bf q}}+\frac h{\overline{\Pi }(T)}\widetilde{{\Pi }}_{y{\bf q}}%
\widetilde{{\Pi }}_{y,-{\bf q}}\right\}  \label{Scl} \\
&&\ \ \ \ \ \ \ \ \ \ +\frac{\overline{{\Pi }}}{3!}\frac{cU}T\sum_{{\bf q}_1%
{\bf q}_2{\bf q}_3}\left( \widetilde{{\bbox \Pi }}_{{\bf q}_1}\widetilde{{%
\bbox \Pi }}_{{\bf q}_2}\right) {\Pi }_{x,{\bf q}_3}\delta ({\bf q}_1+{\bf q}%
_2+{\bf q}_3)+...  \nonumber \\
&&\ \ \ \ \ \ \ \ \ \ +\frac 1{4!}\frac{cU}T\sum_{{\bf q}_1{\bf q}_2{\bf q}_3%
{\bf q}_4}\left( \widetilde{{\bbox \Pi }}_{{\bf q}_1}\widetilde{{\bbox \Pi }}%
_{{\bf q}_2}\right) \left( \widetilde{{\bbox \Pi }}_{{\bf q}_3}\widetilde{{%
\bbox \Pi }}_{{\bf q}_4}\right) \delta ({\bf q}_1+{\bf q}_2+{\bf q}_3+{\bf q}%
_4)+...  \nonumber
\end{eqnarray}
where the field $\widetilde{{\bbox \Pi }}={\bbox \Pi +}(\overline{\Pi },0)$
is now two-component one, and the dots stand for higher-order terms. For the
parameters of the model (\ref{Scl}) we have 
\begin{equation}
R(T)=\frac 13\overline{\Pi }^2(T)U(T)
\end{equation}
and for $d=3$ 
\begin{eqnarray}
\overline{\Pi }^2(T) &=&\frac{18g_R}u\left[ \Theta _0^2-\frac{32\pi ^2}9%
\left( \frac Tc\right) ^2\widetilde{F}_3\left( 0\right) \right]  \nonumber \\
U(T) &=&\frac{8\pi ^2g_R}{\ln (1/\widetilde{h})}\left[ 1+\frac{20\pi ^2g_R}{%
3\ln (1/\widetilde{h})}\widetilde{F}_3^{\prime }\left( \frac{\widetilde{h}%
^2c^2}{4T^2}\right) \right]
\end{eqnarray}
Note that both $R(T)$ and $U(T)$ vanish at $H\rightarrow 0$ as $\ln ^{-1}(%
{\cal I}_0/H)$ due to quantum fluctuations, while $\overline{\Pi }(T)$ is
finite in this limit. The value of $T_c,$ as determined by the condition $%
\overline{\Pi }(T_c)=0,$ read 
\begin{equation}
T_c=\frac{3\sqrt{3}}{2\pi }c\Theta _0  \label{Tc2}
\end{equation}
where $\Theta _0$ is given by (\ref{DD}). Thus the result (\ref{Tc2})
coincides with that obtained in Ref.\cite{Sachdev} up to the nonuniversal
factor $Z_\rho $ As well as for the 3D transverse-field Ising model, in
one-loop order the transition temperature turns out to be proportional to
the ground-state magnetization.

It should be noted that, owing to presence of the gapless Goldstone mode,
the model (\ref{Scl}) is applicable at $T=T_c$ only very close to QPT,
unlike the corresponding model (\ref{Scl1}) for the one-component case. Thus
one can put $Z_{L2}=Z_{L4}=1$. The calculated dependence $T_c(D)$ is shown
and compared with HTSE data in Fig.4. As well as for the one-component case,
the result (\ref{Tc2}) agrees well with HTSE data closely enough to QPT. A
more complete treatment can be performed by considering
quasimomentum-dependent vertices in (\ref{Scl}). This is a complicated task
which is not considered in the present paper.

By the same reason, the model (\ref{Scl}) cannot be used for determining
magnetization below $T_c.$ However, one can expect from the hypothesis of
universality the standard value of the two-component three-dimensional $\phi
^4$ theory critical exponent 
\begin{equation}
\beta _T=\frac 12-\frac{3\epsilon }{2(n+8)}=7/20,
\end{equation}
which is practically the same as in the one-component case ($\beta _T=1/3$).
Unlike the one-component case, the logarithmic correction to $\overline{S}%
(t_{-},T)$ is expected near the temperature phase transition due to the
gapless Goldstone mode.

For $d=2$ the contribution of the gapless mode is logarithmically divergent
and therefore the long-range order at finite temperatures is absent, unlike
the case of the transverse-field Ising model.

\section{Conclusions}

In the present paper we have considered systems that demonstrate in the
ground state a quantum phase transition (QPT). Near QPT the saturation
moment $\overline{S}_0$ is small, but the Curie constant in (\ref{CurieLaw})
is not suppressed. We have $T_c\propto \overline{S}_0$ which is determined
by the value of dynamical critical exponent, $z=1$. The susceptibility (\ref
{HiDis})\ in the intermediate temperature region $\Delta _0\ll T/I\ll 1$ is
determined by the small ground-state energy gap $\Delta _0$ and demonstrates
a $1/T^{d-1}$ behavior. In the strong enough longitudinal magnetic field $%
\Delta _0\ll (H/IS)^{1/3}\ll 1$ the ground state parameters are determined
by magnetic field value rather than by closeness to QPT. The corresponding
dependences have been obtained. Our approach gives a possibility to
investigate both non-universal and universal renormalizations of the
ground-state parameters. The ground-state renormalizations turn out to be
important in the vicinity of QPT. Thus the results for thermodynamic
quantities (e.g. transition temperature) have different forms as functions
of renormalized splitting and bare transverse (external magnetic) field, see
(\ref{tcrg3d}),(\ref{Tc1}) and (\ref{Tc2D-1}),(\ref{Tc2D-2}). This should be
taken into account when treating experimental data.

The discussed class of magnets is similar in some respects to weak itinerant
magnets. Note that for weak itinerant ferromagnets we have $T_c\propto 
\overline{S}_0^{3/2}$ (see, e.g., Ref. \cite{Moriya}), which is due to that
main contribution to thermodynamics comes from paramagnons ($z=3$). As well
as for considered localized-moment systems, calculation of non-universal
ground-state parameters for itinerant magnets is of interest, in particular
for different forms of bare electron density of states.

Now we discuss the experimental situation for some systems exhibiting
magnetic and structural transitions. The compound DyVO$_4$ demonstrates a
structural phase transition at $T_D=14\,$K. The low--lying energy levels in
the spectrum of this system are two Kramers doublets with the splitting $%
\Delta _1=27$cm$^{-1}$ at $T=0$ and $\Delta _2=9$cm$^{-1}$ at $T>T_D$.
Neglecting the Kramers degeneracy one can describe this system by the
transverse-field Ising model with $\Omega /I=1/3$ (see Ref. \cite{Gehring}).
The corresponding point in $\Omega /I$-$T_c$ coordinates is marked in
Fig.1d. This point lies exactly on the HTSE curve and therefore HTSE results
are applicable in this region of $\Omega /I.$ One can see that DyVO$_4$ lies
far from QPT, so that the above-developed theory is not applicable for this
system.

Other systems, which are well described by the transverse-field Ising model,
are the ferroelectric quantum crystals like KH$_2$PO$_4\;$(see, e.g., Ref.%
\cite{Samara}). However, to our knowledge, corresponding detailed data on
ground-state order parameters are absent. To fit experimental data on $T_c$
of Ref. \cite{Samara}, we need explicit dependence of the tunneling
parameter $\Omega $ on pressure.

There are very few experimental data on singlet-singlet systems
demonstrating magnetic phase transitions. The system LiTb$_x$Y$_{1-x}$F$_4$ 
\cite{LiTb} is usually assumed to be characterized by long-range exchange
interactions and therefore well described by the mean-field theory. The
singlet-doublet case is represented by the system NiSi$_2$F$_6$ which is an $%
J=1$ easy-plane Heisenberg magnet. The anisotropy constant is changed under
pressure and thus the value of $D/D_c$ can be varied near unity in the
experiment. The pressure dependence of anisotropy constant was measured
experimentally \cite{FSiNi}. However, to our knowledge, the data on the
pressure dependence of exchange parameter are absent, although it is
supposed to be considerable \cite{HTSE-Pl}. There are also few experimental
data on the ground-state magnetization near QPT. At $p=8.6$Kbar, one has the
experimental values $\overline{J}(T=0)=0.3$ and $T_c=110$mK \cite{FSiNi}.
The calculation according to (\ref{Tc2}) yields ${\cal I}=60$mK which is
larger than the $p=0$ value, ${\cal I}=40$mK\cite{FSiNi:Pars}. Praseodymium
in the dhcp phase contains both ``cubic'' and ``hexagonal'' sites\cite
{CooperBook,Pr}, so that separation of different contributions makes an
additional problem. Generalization of our approach to the singlet-triplet
case in connection with the Pr ions in cubic crystal field will be presented
elsewhere.

Generally, the $1/S$ perturbation theory combined with field-theoretical
scaling analysis enables one to obtain a description of ground-state
properties of transverse-field Ising model, which in a good agreement with
the results of the fourth-order ground-state perturbation theory \cite{GSPT}
for all the values of $\Omega .$ The only fitting parameter used is the
critical field value $\Omega _c.$ The finite-temperature properties are
considered with the use of approach of Ref.\cite{Sachdev}. The same analysis
for the $S=1$ easy-plane Heisenberg model is performed within the expansion
in formal parameter $\lambda (=1)$ which plays the role of $1/S.$ In this
case, besides the critical mode, there is a gapless Goldstone mode, which
considerably modifies the conditions of renormalizations. The consideration
of QPT in degenerate systems induced by the external magnetic field within
the approach used is of interest. In particular, in the case of single-site
anisotropy oscillations of the effective moment with increasing magnetic
field or temperature are expected in such systems with $S>1.$

It is of interest to apply the approach used to various 3D and 2D systems
demonstrating orientational and metamagnetic phase transition with changing
the external magnetic field or anisotropy, e.g. for yttrium garnets\cite
{Levitin} and magnetic films\cite{films}. Depending on a concrete physical
situation, such systems can be described by the strongly anisotropic
transverse-field Ising model or Heisenberg model with small anisotropy.

\appendix

\section{Mapping of the anisotropic Heisenberg model onto the
transverse-field Ising model}

In this Appendix we discuss a possibility of mapping procedure of
anisotropic Heisenberg model (\ref{Hcf}) onto the transverse-field Ising
model. We consider only one important case where the lowest level of $V_{%
\text{cf}}$ is singlet and there is QPT to disordered phase at strong enough 
$V_{\text{cf}}.$ Provided that the first excited state is also singlet,
neglecting all energy levels except lowest and first excited states we can
introduce the pseudospin-$1/2$ operators ${\bf S}${\bf ,} to obtain \cite
{CooperBook,Grover} 
\begin{equation}
V_{\text{cf}}=-\Delta \sum_iS_i^z,\,\,\,\,J_i^z=2\alpha S_i^x  \label{JS}
\end{equation}
where $\alpha =\langle A|J^z|B\rangle $ is the matrix element of ${\bf J,}$ $%
|A\rangle $ and $|B\rangle $ are the lowest and first excited states, $%
\Delta $ is the energy gap between these states. (It should be noted that
left-hand sides of Eqs. (\ref{JS}) act in real-spin space, while right-hand
sides in pseudospin space. Thus the equality signs are used only in the
sense of identity of averages.) Then we obtain the transverse-field Ising
model with $I=4\alpha ^2{\cal I}$ and $\Omega =\Delta .$ The order parameter
of Heisenberg model $\langle J^z\rangle $ is connected with the order
parameter in the transverse-field Ising model by 
\begin{equation}
\langle J^z\rangle =2\alpha \langle S^x\rangle
\end{equation}

Consider now the case where the excited state is a multiplet with the states 
$|B_m\rangle ,$ $m=1...N-1.$ Neglecting the degeneracy of upper energy
level, one can use the same mapping (\ref{JS}) if we choose $\alpha
^2=\sum_{m=1}^{N-1}\langle A|J^z|B_m\rangle ^2.$ However, in this case the
original $SU(N)$ spin space is projected onto $SU(2)$ pseudospin space, and
thus $N-2$ degrees of freedom are neglected. Thus this approach does not
give a possibility to take into account properly the symmetry of the
original model and therefore can be applied only outside the critical
region. To obtain a correct description of such systems in the critical
region one should consider the transition in complete $SU(N)$ space.

The above consideration shows that, in principle, the transverse-field Ising
model (\ref{ITF}) can qualitatively describe singlet magnets even in the
case where the exchange interactions in the true momentum space are
isotropic, as in model (\ref{Hcf}).

\section{Calculation of spin Green's function and order parameter of the
transverse-field Ising model within the $1/S$ expansion}

\label{app:B}

Consider first the disordered phase where $\langle \pi _x\rangle =0.$
Representing $\pi _{zi}=(1+1/S-\pi _{xi}^2-\pi _{yi}^2)^{1/2}$ and assuming $%
\langle \pi _{x,y}^2\rangle \sim 1/S$ (validity of this statement will be
checked below) we expand square root to second order in $1/S$ to obtain 
\begin{eqnarray}
{\cal S}_{\text{dyn}} &=&\frac{iSw_S}2\sum_i\int\limits_0^{1/T}d\tau \left(
\pi _{xi}\frac{\partial \pi _{yi}}{\partial \tau }-\pi _{yi}\frac{\partial
\pi _{xi}}{\partial \tau }\right) +\frac{iS}8\sum_i\int\limits_0^{1/T}d\tau
\left( \pi _{xi}^2+\pi _{yi}^2\right) \left( \pi _{xi}\frac{\partial \pi
_{yi}}{\partial \tau }-\pi _{yi}\frac{\partial \pi _{xi}}{\partial \tau }%
\right)  \label{Ldyn} \\
{\cal S}_{\text{st}} &=&-\frac 12\int\limits_0^{1/T}d\tau \left[
IS^2\sum_{\langle ij\rangle }\pi _{xi}\pi _{xj}+\left( T{\cal P}-\Omega
Sw_S\right) \sum_i\left( \pi _{xi}^2+\pi _{yi}^2\right) -\frac{\Omega S}4%
\sum_i\left( \pi _{xi}^2+\pi _{yi}^2\right) ^2\right]  \label{Lst}
\end{eqnarray}
where $w_S=(1+1/2S)^{-1}$ and ${\cal P}=\sum_{\omega _n}1$ ($\omega _n$
being the Matsubara frequencies) is formally divergent quantity which comes
from the measure of integration, this divergence will be canceled in final
results \cite{Div}. To first order in $1/S$ (we suppose that $\Omega _c\sim
I_0S$) we obtain by standard perturbation theory methods the matrix
two-point vertex function of $\pi _x$,$\,\pi _y$ fields in the form 
\begin{equation}
\Gamma ({\bf q},\omega _n)=\left( 
\begin{array}{cc}
\Omega S\left( w_S+\frac{3X+Y}2\right) +S\Upsilon -T{\cal P}-I_{{\bf q}}S^2
& iS\omega _n\left( w_S+\frac{X+Y}2\right) \\ 
iS\omega _n\left( w_S+\frac{Y+X}2\right) & \Omega S\left( w_S+\frac{3Y+X}2%
\right) +S\Upsilon -T{\cal P}
\end{array}
\right)
\end{equation}
where 
\begin{eqnarray}
X &=&\langle \pi _{xi}^2\rangle =T\sum_{{\bf q},\omega _n}\frac{\Omega /S}{%
\omega _n^2+E_{{\bf q}}^2},  \nonumber \\
\,\,Y &=&\langle \pi _{yi}^2\rangle =T\sum_{{\bf q},\omega _n}\frac{\Omega
/S-I_{{\bf q}}}{\omega _n^2+E_{{\bf q}}^2}  \label{XY}
\end{eqnarray}
and 
\begin{equation}
\,\Upsilon =i\langle \pi _{xi}(\partial \pi _{yi}/\partial \tau )\rangle =%
\frac TS\sum_{{\bf q},\omega _n}\frac{\omega _n^2}{\omega _n^2+E_{{\bf q}}^2}%
,
\end{equation}
with $E_{{\bf q}}=\sqrt{\Omega (\Omega -SI_{{\bf q}})}$ being the bare
``spin-wave'' spectrum in the disordered phase. (One can easily verify that $%
X,Y$ are of the order of $1/S,$ as it was supposed in the beginning). Using
the identity 
\begin{equation}
\Omega SX+S\Upsilon -T{\cal P}=I_0S^2X^{\prime }  \label{Id1}
\end{equation}
where 
\begin{equation}
X^{\prime }=\langle \pi _{xi}\pi _{xj}\rangle =\frac{T\Omega }{I_0S}\sum_{%
{\bf q},\omega _n}\frac{I_{{\bf q}}}{\omega _n^2+E_{{\bf q}}^2}  \label{X'}
\end{equation}
to eliminate the divergences, we derive 
\begin{equation}
\Gamma ({\bf q},\omega _n)=\left( 
\begin{array}{cc}
\Omega S\left( w_S+\frac{X+Y}2\right) +I_0S^2X^{\prime }-I_{{\bf q}}S^2 & 
iS\omega _n\left( w_S+\frac{X+Y}2\right) \\ 
iS\omega _n\left( w_S+\frac{Y+X}2\right) & \Omega S\left( w_S+\frac{3Y-X}2%
\right) +I_0S^2X^{\prime }
\end{array}
\right)  \label{GammaD}
\end{equation}
The value of $\Omega _c$ is determined by the condition $\Gamma _{xx}(0,0)=0$
However, the averages (\ref{XY}) and (\ref{X'}) which determine $1/S$%
-corrections to $\Gamma ({\bf q},\omega _n)$ are $\Omega $-dependent itself.
For consistency, we have to calculate these averages with the zeroth-order
value $\Omega _c=I_0S$. Then we obtain the result for $\Omega _c$ (\ref{Wc})
of main text. At an arbitrary $\Omega $ we perform the replacement $\Omega
\rightarrow I_0S(\Omega /\Omega _c)$ in (\ref{XY}) and (\ref{X'}), which
changes $\Gamma ({\bf q},\omega _n)$ in the order of $1/S^2$ only and gives
a possibility to take into account consistently the shift of $\Omega _c$
owing to quantum fluctuations. Substituting the result for $\Omega _c$ into (%
\ref{GammaD}), we obtain the results (\ref{Gamma}), (\ref{DeltaP}) and (\ref
{DP}) of the main text.

In the ordered phase we rotate the coordinate system through the angle $%
\theta $ around $\pi _y$-axis and again, in new coordinates, expand $%
\widetilde{\pi }_z$ in powers of $\widetilde{\pi }_x,\widetilde{\pi }_y\;$($%
\widetilde{\pi }_y=\pi _y$)$.$ Then we obtain to fourth order 
\begin{eqnarray}
{\cal S}_{\text{st}} &=&-\frac{IS^2}2\sum_{\langle ij\rangle
}\int\limits_0^{1/T}d\tau \left[ -\widetilde{\pi }_{xi}w_S^{-1}\sin 2\theta +%
\widetilde{\pi }_{xi}\widetilde{\pi }_{xj}\cos ^2\theta -(\widetilde{\pi }%
_{xi}^2+\widetilde{\pi }_{xj}^2)\sin ^2\theta \right.  \nonumber \\
&&\ \ \ \ \ \ \ \ \ \ \ \ \ \ \ \left. +\frac 12\widetilde{\pi }_{xj}(%
\widetilde{\pi }_{xi}^2+\widetilde{\pi }_{yi}^2)\sin 2\theta +\frac{(%
\widetilde{\pi }_{xi}^2+\widetilde{\pi }_{yi}^2)(\widetilde{\pi }_{xj}^2+%
\widetilde{\pi }_{yj}^2)-(\widetilde{\pi }_{xi}^2+\widetilde{\pi }_{yi}^2)^2}%
4\sin ^2\theta \right]  \nonumber \\
&&\ \ \ \ \ \ \ \ \ \ \ \ \ \ \ -\Omega S\sum_i\int\limits_0^{1/T}d\tau
\left[ \widetilde{\pi }_{xi}\sin \theta -\frac{\widetilde{\pi }_{xi}^2+%
\widetilde{\pi }_{yi}^2}2w_S\cos \theta -\frac{(\widetilde{\pi }_{xi}^2+%
\widetilde{\pi }_{yi}^2)^2}8\cos \theta \right]  \nonumber \\
&&\ \ \ \ \ \ \ \ \ \ \ \ \ \ \ +\frac{T{\cal P}}2\sum_i\int\limits_0^{1/T}d%
\tau \left( \widetilde{\pi }_{xi}^2+\widetilde{\pi }_{yi}^2\right) ,
\label{SOrd}
\end{eqnarray}
${\cal S}_{\text{dyn}}$ having the same form (\ref{Ldyn}) as in the
disordered phase with the replacement $\pi \rightarrow \widetilde{\pi }$.
Determining the angle $\theta $ from the condition $\langle \widetilde{\pi }%
_x\rangle =0$ we obtain 
\begin{equation}
\cos \theta =\frac \Omega {I_0S}\left[ 1+\frac 1{2S}-\frac{X+2X^{\prime }+Y}2%
\right] ^{-1}  \label{TetITF}
\end{equation}
where 
\begin{eqnarray}
X &=&\langle \widetilde{\pi }_{xi}^2\rangle =T\sum_{{\bf q},\omega _n}\frac{%
I_0}{\omega _n^2+E_{{\bf q}}^2},  \nonumber \\
X^{\prime } &=&\langle \widetilde{\pi }_{xi}\widetilde{\pi }_{xj}\rangle
=T\sum_{{\bf q},\omega _n}\frac{I_{{\bf q}}}{\omega _n^2+E_{{\bf q}}^2}, 
\nonumber \\
\,\,Y &=&\langle \widetilde{\pi }_{yi}^2\rangle =T\sum_{{\bf q},\omega _n}%
\frac{I_0-\eta I_{{\bf q}}}{\omega _n^2+E_{{\bf q}}^2}.  \label{XX'Y}
\end{eqnarray}
$E_{{\bf q}}=S\sqrt{I_0(I_0-\eta I_{{\bf q}})}$ and $\eta =(\Omega /I_0S)^2.$
Besides the averages (\ref{XX'Y}), we introduce the quantity 
\begin{equation}
\,\Upsilon =i\langle \widetilde{\pi }_{xi}(\partial \widetilde{\pi }%
_{yi}/\partial \tau )\rangle =\frac TS\sum_{{\bf q},\omega _n}\frac{\omega
_n^2}{\omega _n^2+E_{{\bf q}}^2}.
\end{equation}
For the two-point vertex function of the fields $\widetilde{\pi }_x,%
\widetilde{\pi }_y$ we have 
\begin{equation}
\Gamma ({\bf q},\omega _n)=\left( 
\begin{array}{cc}
I_0S^2\left( 1+X-\eta X^{\prime }\right) -I_{{\bf q}}S^2W+F_{{\bf q}%
n}+S\Upsilon -T{\cal P} & iS\omega _n\left( w_S+\frac{X+Y}2\right) \\ 
iS\omega _n\left( w_S+\frac{Y+X}2\right) & I_0S^2\left( 1+Y-\eta X^{\prime
}\right) +S\Upsilon -T{\cal P}
\end{array}
\right)
\end{equation}
where 
\begin{equation}
W=\eta (w_S+X+2X^{\prime }+Y)+(1-\eta )X^{\prime },
\end{equation}
The term with 
\begin{eqnarray}
F_{{\bf q}n} &=&-\frac{S^4}{2T}\eta (1-\eta )\sum_{{\bf k},\omega _m}\left[
(2I_{{\bf k}}I_{{\bf k+q}}+4I_{{\bf q}}I_{{\bf k}}+2I_{{\bf k+q}}^2+I_{{\bf q%
}}^2)M_{xxxx}(k,q)\right.  \nonumber \\
&&\ \ \ \ \ \left. +I_{{\bf q}}^2M_{yyyy}(k,q)+2I_{{\bf q}}(I_{{\bf q}}+2I_{%
{\bf k}})M_{xyxy}(k,q)\right]
\end{eqnarray}
where $k=({\bf k,}\omega _m)$, $q=({\bf q,}\omega _n)$ and 
\begin{equation}
M_{\alpha \beta \gamma \delta }(k,q)=\langle \widetilde{\pi }_\alpha (k)%
\widetilde{\pi }_\beta (-k)\rangle \langle \widetilde{\pi }_\gamma (k+q)%
\widetilde{\pi }_\delta (-k-q)\rangle
\end{equation}
($\alpha ,\beta ,\gamma ,\delta =x,y$) arises due to the contribution of the
cubic term in (\ref{SOrd}) in the second order of perturbation theory. This
term has the same order in $1/S$ as other terms and should be retained.
Using the identity 
\begin{equation}
I_0S^2X+S\Upsilon -T{\cal P}=I_0S^2\eta X^{\prime },
\end{equation}
which is an analog of (\ref{Id1}) for the ordered phase, we obtain 
\begin{equation}
\Gamma ({\bf q},\omega _n)=\left( 
\begin{array}{cc}
S^2(I_0-WI_{{\bf q}})+F_{{\bf q}n} & iS\omega _n\left( w_S+X/2+Y/2\right) \\ 
iS\omega _n\left( w_S+X/2+Y/2\right) & I_0S^2\left( 1+Y-X\right)
\end{array}
\right)  \label{GG}
\end{equation}
Performing again the replacement $\Omega \rightarrow I_0S(\Omega /\Omega _c)$
in (\ref{XX'Y}) and reexpressing (\ref{GG}) in terms of $\Omega _c,$ we
obtain the results (\ref{Gamma}), (\ref{DeltaM}) and (\ref{DM}) of the main
text. For the temperature-dependent order parameter we obtain 
\begin{eqnarray}
\overline{S} &\equiv &S\langle \pi _x\rangle =S\sin \theta \langle 
\widetilde{\pi }_x\rangle  \nonumber \\
\ &=&S\left\{ 1-\eta -\frac \eta {2S}\sum_{{\bf q}}\frac{2I_0+\eta I_{{\bf q}%
}}{\sqrt{I_0(I_0-\eta I_{{\bf q}})}}\coth \frac{S\sqrt{I_0(I_0-\eta I_{{\bf q%
}})}}{2T}\right.  \nonumber \\
&&\ \ \ \ \ \ \ \ \left. +\frac \eta {2S}\sum_{{\bf q}}\frac{2I_0+I_{{\bf q}}%
}{\sqrt{I_0(I_0-I_{{\bf q}})}}\right\} ^{1/2}\left( 1+\frac 1{2S}-\frac{X+Y}2%
\right)  \label{Magn}
\end{eqnarray}
Note that near QPT the last multiplier in this expression is close to unity
and only slightly temperature-dependent. Thus it can be replaced by its
zero-temperature value.

\section{Rotation in $SU(3)$ space for the easy-plane Heisenberg model}

\label{app:C}

Following to Refs. \cite{Fl-Epl,Onufr,Valkov} we perform in the Hamiltonian (%
\ref{He-pl}) the unitary transformation 
\begin{equation}
\widetilde{X}_i^{pq}=U^{\dagger }(\theta )X_i^{pq}U(\theta )
\end{equation}
with 
\begin{eqnarray}
U(\theta ) &=&\exp [\theta (X^{-1,1}-X^{1,-1})/2]  \nonumber \\
\ &=&1+[\cos (\theta /2)-1](X^{-1,-1}+X^{1,1})+\sin (\theta
/2)(X^{-1,1}-X^{1,-1})
\end{eqnarray}
Then the Hamiltonian takes the form 
\begin{equation}
{\cal H}={\cal H}^{(1)}+{\cal H}^{(2)}
\end{equation}
with 
\begin{eqnarray}
{\cal H}^{(1)} &=&\frac 12\sum_i\left[ (D\sin \theta +2{\cal I}_0\cos
^2\theta )(2X_i^{11}+X_i^{00})+DX_i^{00}+(D-2{\cal I}_0\sin \theta )\cos
\theta (X_i^{1,-1}+X_i^{-1,1})\right]  \nonumber \\
{\cal H}^{(2)} &=&-\frac{{\cal I}}2\sum_{<ij>}\left[ \cos \theta
(2X_i^{11}+X_i^{00})-\sin \theta (X_i^{1,-1}+X_i^{-1,1})\right] \left[ \cos
\theta (2X_j^{11}+X_j^{00})-\sin \theta (X_j^{1,-1}+X_j^{-1,1})\right] 
\nonumber \\
&&\ \ \ \ \ \ \ \ \ \ \ \ \ \ \ \ \ \ \ \ \ \ \ \ -\frac{{\cal I}}2%
\sum_{<ij>}\sum_{\sigma =\pm 1}\left[ 2X_i^{0\sigma }X_j^{\sigma 0}+\sigma
\sin \theta (X_i^{0\sigma }X_j^{0\sigma }+X_i^{\sigma 0}X_j^{\sigma 0})+\cos
\theta (X^{-\sigma 0}X^{\sigma 0}+X^{0,-\sigma }X^{0\sigma })\right]
\end{eqnarray}
where ${\cal I}_0=2d{\cal I}$ and we have dropped the tilde sign at the $X$%
-operators$.$ Further we represent Hubbard operators via boson ones\cite
{Onufr,Valkov} 
\begin{eqnarray}
X^{1,-1} &=&a^{\dagger }(1-\lambda a^{\dagger }a-\lambda b^{\dagger }b)^{1/2}
\nonumber \\
X^{0,-1} &=&b^{\dagger }(1-\lambda a^{\dagger }a-\lambda b^{\dagger }b)^{1/2}
\nonumber \\
X^{10} &=&a^{\dagger }b,\,\,\,X^{00}=b^{\dagger }b,\,\,\,X^{11}=a^{\dagger }a
\label{HubBose1}
\end{eqnarray}
where $\lambda (=1)$ is the parameter introduced to construct perturbation
theory (cf. the Holstein-Primakoff expansion in the case of a Heisenberg
magnet). Then we obtain the Hamiltonian of the bosons 
\begin{equation}
{\cal H}={\cal H}_1+{\cal H}_2+{\cal H}_3+{\cal H}_4+...  \label{H1234}
\end{equation}
where 
\begin{eqnarray}
{\cal H}_1 &=&\frac 12\cos \theta (D-2{\cal I}_0\sin \theta )\sum_i\left(
a_i^{\dagger }+a_i\right) \\
{\cal H}_2 &=&(D\sin \theta +2{\cal I}_0\cos ^2\theta )\sum_ia_i^{\dagger
}a_i+[D(1+\sin \theta )/2+{\cal I}_0\cos ^2\theta ]\sum_ib_i^{\dagger }b_i 
\nonumber \\
&&\ \ \ \ \ \ \ \ -\frac{{\cal I}}2\sum_{\langle ij\rangle }\left[ \sin
^2\theta \left( a_i^{\dagger }+a_i\right) \left( a_j^{\dagger }+a_j\right)
+2b_i^{\dagger }b_j-\sin \theta \left( b_i^{\dagger }b_j^{\dagger
}+b_ib_j\right) \right] \\
{\cal H}_3 &=&-\frac 12\cos \theta (D-2{\cal I}_0\sin \theta )\sum_i\left[
a_i^{\dagger }\left( a_i^{\dagger }a_i+b_i^{\dagger }b_i\right) +\left(
a_i^{\dagger }a_i+b_i^{\dagger }b_i\right) a_i\right]  \nonumber \\
&&\ \ \ \ \ \ \ \ +{\cal I}\cos \theta \sum_{\langle ij\rangle }\left[ \sin
\theta \left( 2a_i^{\dagger }a_i+b_i^{\dagger }b_i\right) \left(
a_j^{\dagger }+a_j\right) -\left( b_ia_j^{\dagger }b_j+b_i^{\dagger
}b_j^{\dagger }a_j\right) \right] \\
{\cal H}_4 &=&-\frac{{\cal I}}2\sum_{\langle ij\rangle }\left[ \sin ^2\theta
\left( a_i^{\dagger }a_i+2b_i^{\dagger }b_i\right) \left( a_j^{\dagger
}a_j+2b_j^{\dagger }b_j\right) +2a_i^{\dagger }a_jb_j^{\dagger }b_i\right. 
\nonumber \\
&&\ \ \ \left. -\cos ^2\theta \left( b_i^{\dagger }+b_i\right) \left(
a_j^{\dagger }a_jb_j^{\dagger }+b_j^{\dagger }b_j^{\dagger }b_j+\text{h.c.}%
\right) +\cos 2\theta \left( a_ia_jb_i^{\dagger }b_j^{\dagger }+\text{h.c.}%
\right) \right.  \nonumber \\
&&\ \ \ \left. -\left( a_i^{\dagger }a_j^{\dagger }a_ja_j+a_i^{\dagger
}b_j^{\dagger }b_ja_j+\text{h.c.}\right) +\cos 2\theta \left( a_i^{\dagger
}a_i^{\dagger }a_ia_j^{\dagger }+a_i^{\dagger }b_i^{\dagger }b_ia_j^{\dagger
}+\text{h.c.}\right) \right]
\end{eqnarray}
and we have omitted terms containing more than four Bose operators.
Diagonalizing the quadratic part ${\cal H}_2$ of the Hamiltonian (\ref{H1234}%
) we obtain 
\begin{equation}
{\cal H}_2=\sum_{{\bf k}}\left( E_{{\bf k}\alpha }\alpha _{{\bf k}}^{\dagger
}\alpha _{{\bf k}}+E_{{\bf k}\beta }\beta _{{\bf k}}^{\dagger }\beta _{{\bf k%
}}\right)
\end{equation}
where the spectra of excitations are given by 
\begin{eqnarray}
E_{{\bf k}\alpha } &=&2\sqrt{I_0(I_0-\eta ^2I_{{\bf k}})},  \label{E-plEka}
\\
E_{{\bf k}\beta } &=&\sqrt{(1+\eta )(I_0-I_{{\bf k}})[I_0-I_{{\bf k}}+\eta
(I_0+I_{{\bf k}})]},  \label{E-plEkb}
\end{eqnarray}
$\eta =\sin \theta =D/D_c$. The angle $\theta $ is determined by the
condition $\langle a_i\rangle =0.$ To first order in the formal parameter $%
\lambda $ we have 
\begin{eqnarray}
\sin \theta &=&\frac D{2I_0}\left[ 1-\lambda \left( R_1+R_2\right) \right]
^{-1}  \label{E-plTet} \\
R_1(T) &=&2\sum_{{\bf k}}\frac{2I_0+(2-\eta ^2)I_{{\bf k}}}{2E_{{\bf k}%
\alpha }}(1+2N_{{\bf k}\beta })-1  \nonumber \\
R_2(T) &=&\sum_{{\bf k}}\frac{(1+\eta )I_0-I_{{\bf k}}+\eta I_{{\bf k}}^2/I_0%
}{2E_{{\bf k}\beta }}(1+2N_{{\bf k}\alpha })-\frac 12  \label{R1R2}
\end{eqnarray}
The gap in the excitation spectrum for the mode $\alpha $ to first order in $%
\lambda $ is given by 
\begin{eqnarray}
\widetilde{E}_{{\bf k}=0,\alpha } &=&2\sqrt{(1+B_0+B_1)}{\cal I}_0\Delta _{-}
\nonumber \\
\Delta _{-}^2 &=&(1+A_0+A_1)(1-\eta ^2)
\end{eqnarray}
where 
\begin{eqnarray}
A_0 &=&\sum_{{\bf k}}\left[ \frac{\eta ^2\gamma _{{\bf k}}-4\gamma _{{\bf k}%
}-4}{E_{{\bf k}\alpha }}+\frac{\gamma _{{\bf k}}(1+\eta +\eta ^2)/(1+\eta
)-(\eta +1)}{E_{{\bf k}\beta }}\right]  \label{A0} \\
B_0 &=&-\frac 12\sum_{{\bf k}}\left[ \frac{8(1+\gamma _{{\bf k}})-20\eta
^2\gamma _{{\bf k}}+11\eta ^4\gamma _{{\bf k}}}{E_{{\bf k}\alpha }}+\frac{%
2(1+\eta )-2\gamma _{{\bf k}}+2\eta ^3\gamma _{{\bf k}}^2}{E_{{\bf k}\beta }}%
\right] \\
A_1 &=&2\eta ^2\sum_{{\bf k}}\frac{(5/2)(2-\eta ^2\gamma _{{\bf k}%
})^2+4(2-\eta ^2\gamma _{{\bf k}})\eta ^2\gamma _{{\bf k}}+(5/2)(\eta
^2\gamma _{{\bf k}})^2}{E_{{\bf k}\alpha }^3}  \nonumber \\
&&\ \ \ \ \ \ +\frac 12\sum_{{\bf k}}\frac{(\eta ^2+1)(1+\eta -\gamma _{{\bf %
k}})^2+(\eta ^2+1)(\eta \gamma _{{\bf k}})^2+4\eta (1+\eta -\gamma _{{\bf k}%
})(\eta \gamma _{{\bf k}})}{E_{{\bf k}\beta }^3}  \label{A1} \\
B_1 &=&\frac 12(1-\eta ^2)\sum_{{\bf k}}\left[ \frac{4\eta ^2(2-\eta
^2\gamma _{{\bf k}})}{E_{{\bf k}\alpha }^3}+\frac{(1+\eta -\gamma _{{\bf k}%
})^2-(\eta \gamma _{{\bf k}})^2}{E_{{\bf k}\beta }^3}\right]
\end{eqnarray}
The mode $\beta $ has Goldstone type and is gapless to arbitrary order in $%
\lambda $.

\section{Renormalization of the two-component $\phi ^4$ model with
spontaneously broken symmetry}

\label{app:D}

In this Appendix we consider the renormalization of the action (\ref{S2}) in
the ordered phase, $m^2<0.$ Introducing the quantity $\kappa ^2=-2m^2>0$ and
performing the shift $\pi _x\rightarrow \pi _x+\pi _0$ we obtain 
\begin{eqnarray}
{\cal S} &=&\frac 12\int\limits_0^{c/T}d\tau \int d^dr\left[ \,(\partial %
\bbox{\pi })^2+(\kappa ^2+3\widetilde{h})\pi _x^2+\widetilde{h}\pi
_y^2\right]  \nonumber \\
&&\ \ \ \ \ \ \ \ \ \ \ \ +\frac u{4!}\int\limits_0^{c/T}d\tau \int
d^dr\,\left( 4\pi _x+\bbox{\pi }^2\right) \bbox{\pi }^2  \label{S1}
\end{eqnarray}
where $\widetilde{h}=h/\overline{\pi }_0$, $\overline{\pi }_0=(3\kappa
^2/u)^{1/2}$ and $\pi _0=\overline{\pi }_0(1+\widetilde{h}/\kappa ^2)$ is
determined by the requirement of absence in the action of terms that are
linear in $\pi _x$. In these notations the condition of smallness of
magnetic field is $\widetilde{h}^{1/2}\ll \kappa $ (the condition of
closeness to QPT $\kappa \ll \Lambda $ is also assumed). Under these
conditions, the action (\ref{S1}) has two characteristic lengths, $1/%
\widetilde{h}^{1/2}$ and $1/\kappa $. This situation is the same as in the
theory of crossover phenomena \cite{Cross}. Further we follow to Ref.\cite
{Cross} to include exactly the smaller characteristic length into $Z$%
-factors. Then we obtain to one-loop order 
\begin{eqnarray}
Z^{\text{cont}} &=&1+{\cal O}(g^2),  \nonumber \\
\;Z_2^{\text{cont}} &=&1+\frac g{2\varepsilon }\frac 1{(1+\kappa ^2/\mu
^2)^{\varepsilon /2}}+\frac g{6\varepsilon }-\frac{2g}{3\varepsilon }\frac{%
\mu ^\varepsilon }{\Lambda ^\varepsilon },  \nonumber \\
\;Z_4^{\text{cont}} &=&1+\frac{3g}{2\varepsilon }\frac 1{(1+\kappa ^2/\mu
^2)^{\varepsilon /2}}+\frac g{6\varepsilon }-\frac{5g}{3\varepsilon }\frac{%
\mu ^\varepsilon }{\Lambda ^\varepsilon }.
\end{eqnarray}
(note that the Ward identities guarantee that the structure of the
interaction term is preserved by renormalizations, and one renormalization
constant is sufficient to renormalize all four-particle vertex functions,
see, e.g., Ref.\cite{Amit}). The flow functions for the
effective-Hamiltonian parameters are 
\begin{eqnarray}
\beta (g,\kappa /\mu ) &=&-\varepsilon g+\frac{3g^2}2\frac 1{1+\kappa ^2/\mu
^2}+\frac{g^2}6  \nonumber \\
\gamma (g,\kappa /\mu ) &=&-\frac g2\frac 1{1+\kappa ^2/\mu ^2}-\frac g6
\end{eqnarray}
Putting in these expressions $\varepsilon =0$, performing the integration
and supposing that scaling starts at $\mu \gg \kappa $ we obtain the
effective-Hamiltonian parameters at the scale $\mu \rho $ 
\begin{eqnarray}
\frac 1{g_\rho } &=&\frac 1g-\frac 34\ln (\rho ^2+\kappa ^2/\mu ^2)-\frac 16%
\ln \rho  \nonumber \\
\kappa _\rho ^2 &=&\kappa ^2\exp \left[ \frac g2\int\limits_1^\rho d\rho
^{\prime }\left( \frac{\rho ^{\prime }}{\rho ^{\prime 2}+\kappa ^2/\mu ^2}+%
\frac 1{3\rho ^{\prime }}\right) \right.  \nonumber \\
&&\ \ \ \ \ \ \ \ \ \ \ \ \left. \times \frac 1{1-(g/6)\ln \rho ^{\prime
}-(3g/4)\ln (\rho ^{\prime 2}+\kappa ^2/\mu ^2)}\right]  \label{gk2}
\end{eqnarray}
Our plan now is to use these scaling formulas to reach the scale $\mu \rho
\sim \widetilde{h}^{1/2}\ll \kappa .$ For these values of $\rho $ the
formulas (\ref{gk2}) are simplified: 
\begin{eqnarray}
g_\rho ^{-1} &=&g^{-1}\left[ 1-(3g/2)\ln (\kappa /\mu )-(g/6)\ln \rho \right]
\nonumber \\
\kappa _\rho ^{-2} &=&\kappa ^{-2}\left[ 1-(3g/2)\ln (\kappa /\mu )-(g/6)\ln
\rho \right]  \nonumber \\
&&\ \ \ \ \ \ \ \ \ \ \ \ \times \left[ 1-(5g/3)\ln (\kappa /\mu )\right]
^{-3/5}\Phi _0(g,\kappa ^2/\mu ^2)  \label{gt}
\end{eqnarray}
where $\Phi _0(g,x)$ is given by 
\begin{eqnarray}
\ln \Phi _0(g,x) &=&\frac g{12}\int\limits_0^1\frac{dy}y\left[ \frac{4y+x}{%
y+x}\frac 1{1-(g/12)\ln y-(3g/4)\ln (x+y)}\right.  \nonumber \\
&&\ \ \ \ \ \ \ \ \ \ \ \ \ \ \left. -\frac{\theta (x-y)}{1-(g/12)\ln
y-(3g/4)\ln x}-\frac{4\theta (y-x)}{1-(5g/6)\ln y}\right]  \label{F0}
\end{eqnarray}
and $\theta (x)$ is the step function. At extremely small $x$ we have $\Phi
_0(g,x)\simeq \exp (1/\ln ^2x)\simeq 1$. The result (\ref{gt}) with $\Phi
_0(g,\kappa ^2/\mu ^2)=1$ can be obtained more directly if we perform the
scaling procedure in two steps: at the first step $\rho \gg \kappa /\mu $
and the flow functions are the same as for the two-component isotropic $\phi
^4$ model, while at the second step $\rho \ll \kappa /\mu $ and the flow
functions include only contributions of the Goldstone modes. The scaling
formulas are joined at $\rho =\kappa /\mu $. However, this procedure does
not give a possibility to describe correctly the contribution of the
crossover region $\rho \sim \kappa /\mu $. Putting in the above results $\mu
=\Lambda =(2d)^{1/2},$ we obtain the result (\ref{scal2}) of the main text.

At finite but small $\varepsilon $ we obtain in a similar way 
\begin{eqnarray}
g_\rho ^{-1} &=&g^{-1}\left[ 1+(3g/2\varepsilon )(\kappa ^{-\varepsilon
}/\mu ^{-\varepsilon }-1)+(g/6\varepsilon )(\rho ^{-\varepsilon }-1)\right]
\\
\kappa _\rho ^{-2} &=&\kappa ^{-2}\left[ 1+(3g/2\varepsilon )(\kappa
^{-\varepsilon }/\mu ^{-\varepsilon }-1)+(g/6\varepsilon )(\rho
^{-\varepsilon }-1)\right]  \nonumber \\
&&\ \ \ \ \ \ \ \ \ \ \ \ \times \left[ 1-(5g/3\varepsilon )(\kappa
^{-\varepsilon }/\mu ^{-\varepsilon }-1)\right] ^{-3/5}\Phi _\varepsilon
(g,\kappa ^2/\mu ^2)
\end{eqnarray}
with some function $\Phi _\varepsilon (g,x)\simeq 1.$

{\sc Figure captions}

Fig.1. Ground-state energy gap $\widetilde{E}_0(\Omega )$ ($\Omega >\Omega
_c,$ a) and order parameter $\overline{S}(\Omega )$ ($\Omega <\Omega _c$, b)
for the 3D transverse-field Ising model in different approaches. The value $%
\Omega _c=2.58I$ is used for calculating the $1/S^{\prime }$ and RG$^{\prime
}$ curves. Arrow shows the value of the critical field $\Omega _c,$ obtained
by $1/S$-expansion

Fig.2. Ground-state energy gap $\widetilde{E}_0(\Omega )$ ($\Omega >\Omega
_c,$a) and order parameter $\overline{S}(\Omega )$ ($\Omega <\Omega _c$, b)
for the 2D transverse-field Ising model in different approaches. The
notations used are the same as in Fig.1

Fig.3. Transition temperature as a function of $\Omega /I_0$ for the 3D
transverse-field Ising model in different approaches.

Fig.4. Transition temperature as a function of $D/{\cal I}_0$ for an
easy-plane ferromagnet in different approaches


\begin{references}
\bibitem{Div}  One can obtain the same results in a more correct way (which
is free from divergences) by keeping along the calculation the imaginary
time to be discrete, $\tau =\tau _i=i/{\cal N}T\;$with $i=0...{\cal N}$, and
passing to the limit ${\cal N}\rightarrow \infty $ only at the end.
\end{references}
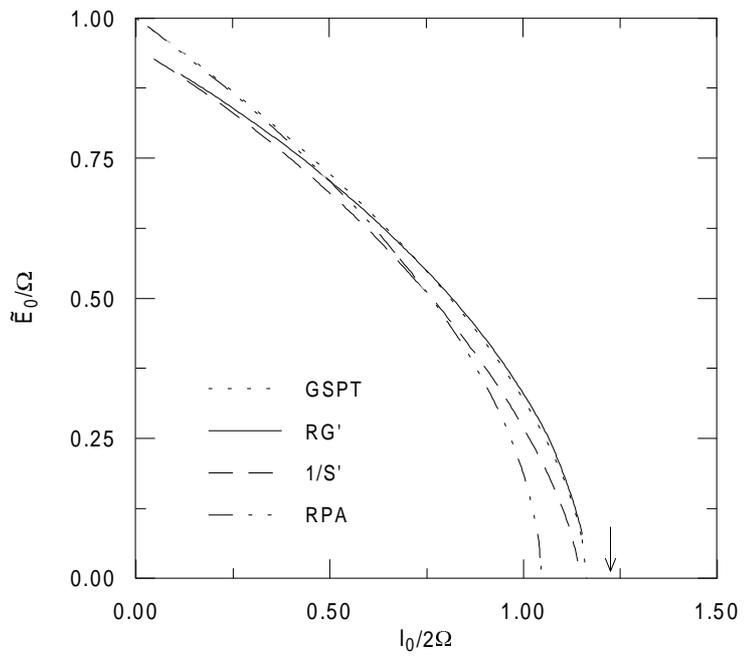
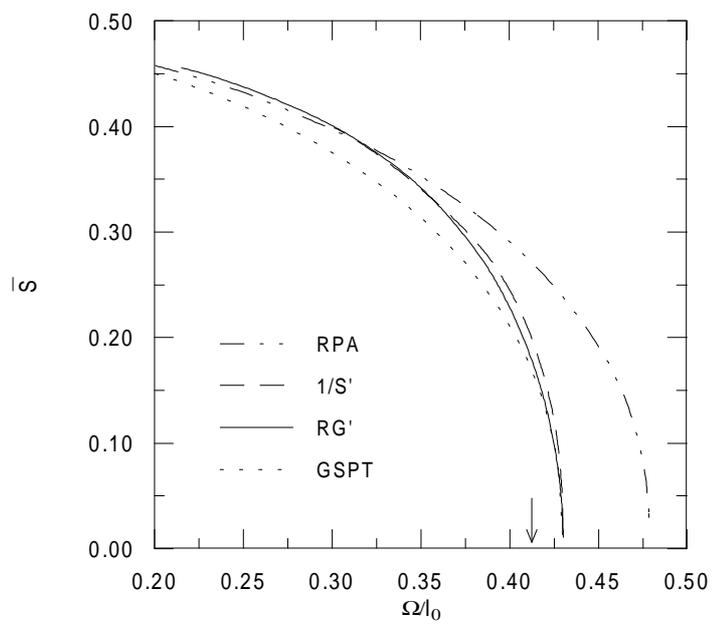
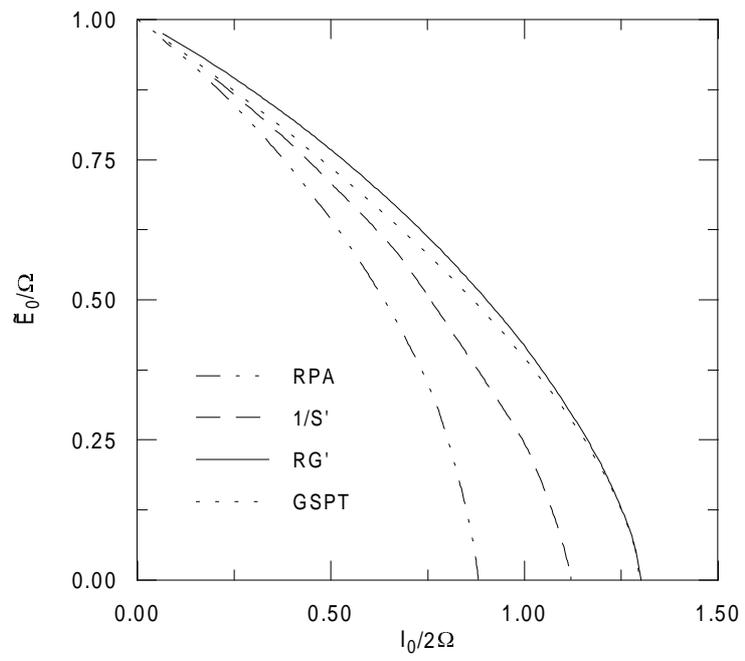
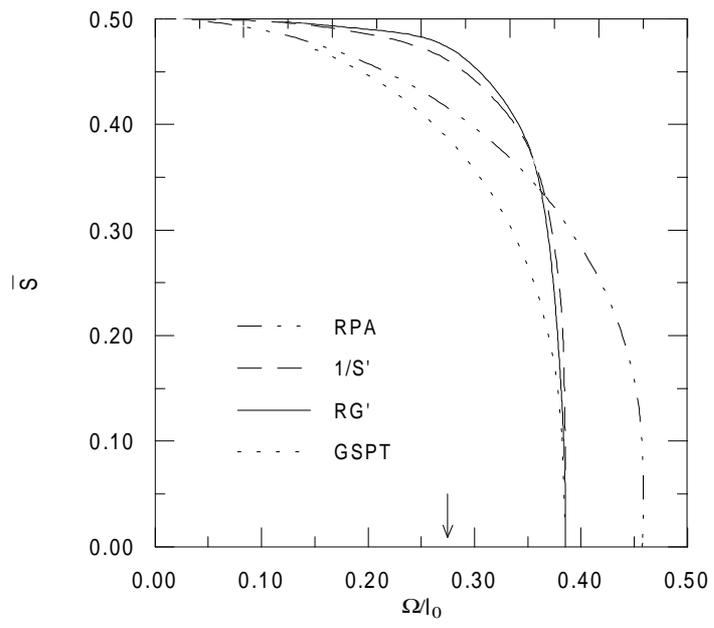
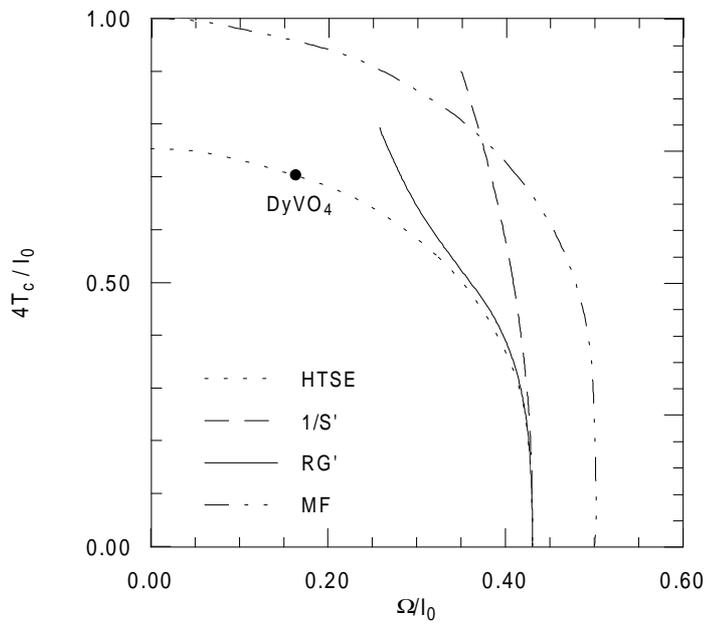
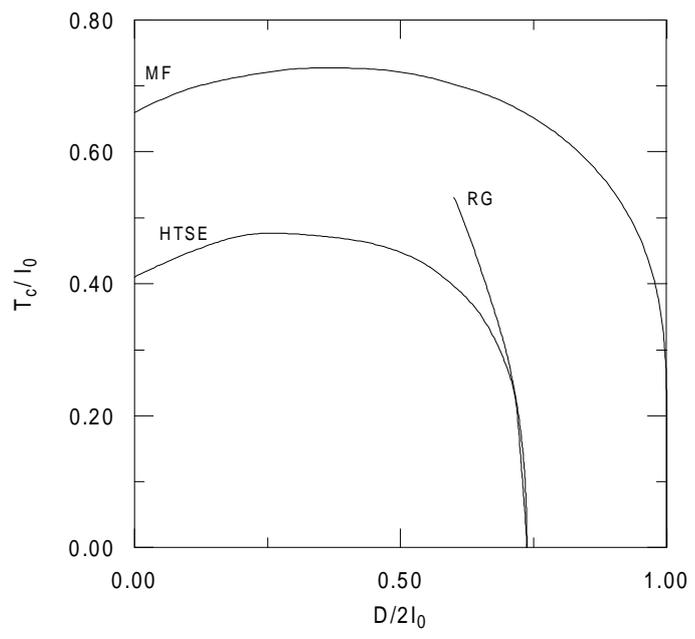
\end{document}